\documentclass[twocolumn]{aastex631}

\usepackage{graphicx}
\usepackage{amsmath} 
\usepackage{amssymb} 
\usepackage{latexsym}

\newcommand{\MAXI}{MAXI J0556--332}

\newcommand{\be}{\begin{equation}}
\newcommand{\ee}{\end{equation}}
\newcommand{\ba}{\begin{eqnarray}}
\newcommand{\ea}{\end{eqnarray}}

\newcommand{\Eq}[1]{Equation~(\ref{#1})}
\newcommand{\Fig}[1]{Figure~\ref{#1}}
\newcommand{\Tab}[1]{Table~\ref{#1}}
\newcommand{\Sec}[1]{Section~\ref{#1}}
\newcommand{\App}[1]{Appendix~\ref{#1}}

\shorttitle{A Hyperburst in the MAXI J0556--332 Neutron Star}
\shortauthors{Page et al.}

\begin{document}
%%%%%%%%%%%%%%%%%%%%%%%%%%%%%%%%%%%%%%%%%%%%%%%%%%%%%%%%%%%%%%%%%%%%%%%%%%%%%%%%
%%%%%%%%%%%%%%%%%%%%%%%%%%%%%%%%%%%%%%%%%%%%%%%%%%%%%%%%%%%%%%%%%%%%%%%%%%%%%%%%
%%%%%%%%%%%%%%%%%%%%%%%%%%%%%%%%%%%%%%%%%%%%%%%%%%%%%%%%%%%%%%%%%%%%%%%%%%%%%%%%
%%%%%%%%%%%%%%%%%%%%%%%%%%%%%%%%%%%%%%%%%%%%%%%%%%%%%%%%%%%%%%%%%%%%%%%%%%%%%%%%
%%%%%%%%%%%%%%%%%%%%%%%%%%%%%%%%%%%%%%%%%%%%%%%%%%%%%%%%%%%%%%%%%%%%%%%%%%%%%%%%

\title{A ``Hyperburst'' in the MAXI J0556-332 Neutron Star: \\
Evidence for a New Type of Thermonuclear Explosion}

\author[0000-0003-2498-4326]{Dany Page}
\affiliation{Instituto de Astronom\'ia, Universidad Nacional Aut\'onoma de M\'exico, Ciudad de  M\'exico, CDMX 04510, Mexico}
\email{page@astro.unam.mx}

\author[0000-0001-8371-2713]{Jeroen Homan}
\affiliation{Eureka Scientific, Inc., 2452 Delmer Street, Oakland, CA 94602, USA}
\email{jeroenhoman@icloud.com}

\author[0000-0003-2334-6947]{Martin Nava-Callejas}
\affiliation{Instituto de Astronom\'ia, Universidad Nacional Aut\'onoma de M\'exico, Ciudad de  M\'exico, CDMX 04510, Mexico}
\email{mnava@astro.unam.mx}

\author[0000-0002-6447-3603]{Yuri Cavecchi}
\affiliation{Instituto de Astronom\'ia, Universidad Nacional Aut\'onoma de M\'exico, Ciudad de  M\'exico, CDMX 04510, Mexico}
\email{ycavecchi@astro.unam.mx}

\author[0000-0002-7326-7270]{Mikhail V. Beznogov}
\affiliation{National Institute for Physics and Nuclear Engineering (IFIN-HH), RO-077125 Bucharest, Romania}
\email{mikhail.beznogov@nipne.ro}

\author[0000-0002-0092-3548]{Nathalie Degenaar}
\affiliation{Anton Pannekoek Institute for Astronomy, University of Amsterdam, Postbus 94249, 1090 GE Amsterdam, The Netherlands}
\email{degenaar@uva.nl}

\author[0000-0002-3516-2152]{Rudy Wijnands}
\affiliation{Anton Pannekoek Institute for Astronomy, University of Amsterdam, Postbus 94249, 1090 GE Amsterdam, The Netherlands}
\email{r.a.d.wijnands@uva.nl}

\author{Aastha S. Parikh}
\affiliation{Anton Pannekoek Institute for Astronomy, University of Amsterdam, Postbus 94249, 1090 GE Amsterdam, The Netherlands}
%\email{a.s.parikh@uva.nl}

%%%%%%%%%%%%%%%%%%%%%%%%%%%%%%%%%%%%%%%%%%%%%%%%%%%%%%%%%%%%%%%%%%%%%%%%%%%%%%%%
\begin{abstract}
%%%%%%%%%%%%%%%%%%%%%%%%%%%%%%%%%%%%%%%%%%%%%%%%%%%%%%%%%%%%%%%%%%%%%%%%%%%%%%%%

The study of transiently accreting neutron stars provides a powerful means to elucidate the properties of neutron star crusts.
We present extensive  numerical simulations of the evolution of the neutron star in the transient low-mass X-ray binary MAXI J0556--332. 
We model nearly twenty observations obtained during the quiescence phases after four different outbursts of the source in the past decade, considering the heating of the star during accretion by the deep crustal heating mechanism complemented by some shallow heating source.
We show that cooling data are consistent with a single source of shallow heating acting during the last three outbursts, while a very different and powerful energy source is required to explain the extremely high effective temperature of the neutron star, $\sim 350$ eV, when it exited the first observed outburst.
We propose that a gigantic thermonuclear explosion, a ``hyperburst'' from unstable burning of neutron rich isotopes of oxygen or neon, occurred a few weeks before the end of the first outburst, releasing $\sim10^{44}$ ergs at densities of the order of $10^{11}$ g cm$^{-3}$.
This would be the first observation of a hyperburst and these would be extremely rare events as the build up of the exploding layer requires about a millennium of accretion history.
Despite its large energy output, the hyperburst did not produce, due to its depth, any noticeable increase in luminosity during the accretion phase and is only identifiable by its imprint on the later cooling of the neutron star.
 
%%%%%%%%%%%%%%%%%%%%%%%%%%%%%%%%%%%%%%%%%%%%%%%%%%%%%%%%%%%%%%%%%%%%%%%%%%%%%%%%
\end{abstract}
%%%%%%%%%%%%%%%%%%%%%%%%%%%%%%%%%%%%%%%%%%%%%%%%%%%%%%%%%%%%%%%%%%%%%%%%%%%%%%%%


%%%%%%%%%%%%%%%%%%%%%%%%%%%%%%%%%%%%%%%%%%%%%%%%%%%%%%%%%%%%%%%%%%%%%%%%%%%%%%%%
\section{Introduction} 
\label{Sec:Intro}
%%%%%%%%%%%%%%%%%%%%%%%%%%%%%%%%%%%%%%%%%%%%%%%%%%%%%%%%%%%%%%%%%%%%%%%%%%%%%%%%

\begin{figure}
	\begin{center}
	\includegraphics[width=0.99\columnwidth]{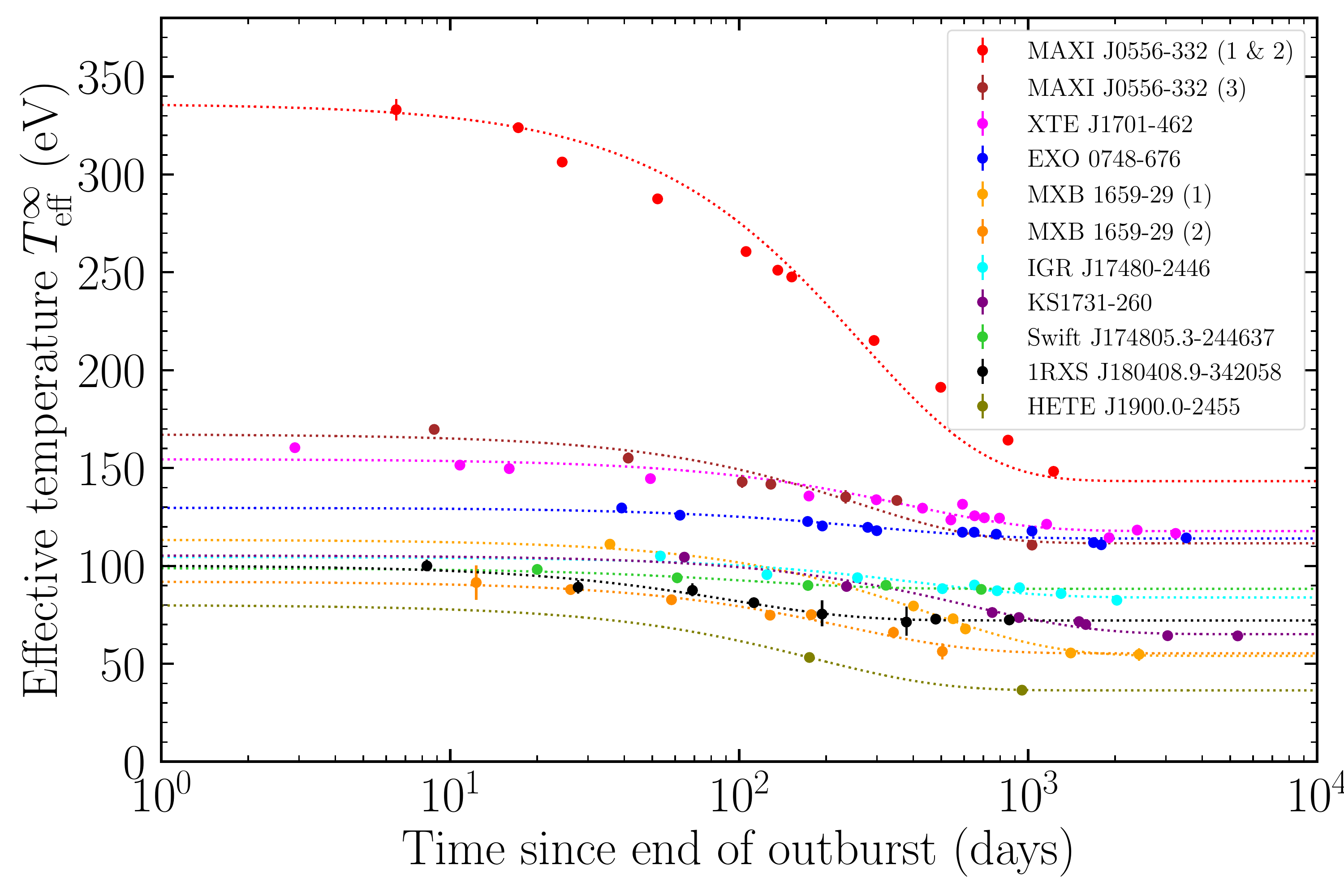}
	\end{center}
	\caption{
	Present sample of cooling observations after long accretion outbursts from the sources:
	MAXI J0556-332 outbursts 1 \& 2 together
	(outburst 2 was small and had almost no effect on the cooling curve: see \Fig{fig:Bigbadaboom} below)
	and 3 (\citealt{Parikh:2017uj} and this work),
	XTE J1701-462 \citep{Parikh:2020ue},
	EXO 0748-676 \citep{Parikh:2020ue},
	two outbursts from
	MXB 1659-29 \citep{Parikh:2019aa},
	IGR J17480-2446 \citep{Ootes:2019aa},
	KS1731-260 \citep{Merritt:2016uu},
	Swift J174805.3-244637 \citep{Degenaar:2015aa},
	1RXS J180408.9-342058 \citep{Parikh:2017aa,Parikh:2018tg},
	and HETE J1900.0-2455 \citep{Degenaar:2021us}.
	Dotted lines show simple fits of the form $T(t) = T_0 + \Delta T \exp(-t/\tau)$ to guide the eye and have no claim to be physically meaningful.
	}
	\label{fig:Cool_All}
\end{figure}

Observations of the cooling of neutron stars in transient low-mass X-ray binaries
after a long phase of accretion have opened a new window in the study of neutron star interiors.
During accretion, compression of matter in the neutron star crust induces a series of non-equilibrium reactions,
such as electron captures, neutron emissions, and pycnonuclear fusions \citep{Sato:1979aa,Bisnovatyi-Kogan:1979aa,Haensel:1990kx}.
The energy generated by these reactions slowly diffuses into the neutron star core, a mechanism known as ``deep crustal heating'' \citep{Brown:1998aa}
which, over a long time, will lead to an equilibrium between this heating and  photon and neutrino cooling mechanisms \citep{Miralda-Escude:1990uu,Brown:1998aa,Colpi:2001vn}.
Depending on the star's core temperature, the timescale to establish this equilibrium can range from a few decades for an initially very cold core, up to millions of years
for the hottest stars \citep{Wijnands:2013up}.
Regarding evolution on short timescales, theoretical modeling found that in the case of a long and strong enough accretion outburst the crust can be driven 
out of thermal equilibrium with the core \citep{Rutledge:2002tu}, which led to the prediction that subsequent cooling of the crust, 
once accretion has stopped and the surface temperature is  not anymore controlled by the mass accretion but rather by the internal evolution of the star, 
should be observable on a timescale of a few years.
This prediction has been amply confirmed and to date crust cooling after an accretion outburst has been observed in almost a dozen cases,
which we reproduce in \Fig{fig:Cool_All}.

%------------------------------------------------------------------------------------------------
%-----------------------------------------------------------------------------------------------

After observations of the cooling of the first quasi-persistent source that went into quiescence 
(KS 1732-260: \citealt{Wijnands:2001wp,Cackett:2006wb}), theoretical modeling found it was not possible to
reproduce the high temperature of the first data point, obtained two months after the end of the outburst, within the deep crustal heating scenario \citep{Shternin:2007aa}
and the introduction of another energy source, located at low densities and dubbed ``shallow heating'',  was found to be necessary in the models \citep{Brown:2009aa}.
The deep crustal heating generates an energy $Q_\mathrm{dc}$ of about 1.5 to 2 MeV per accreted nucleon \citep{Gupta:2008aa,Haensel:2008aa,Fantina:2018ub,Shchechilin:2021wi},
most of it through pycnonuclear fusions in the inner crust at densities above $10^{12}$ g cm$^{-3}$.
In contrast, shallow heating has been found to deposit energy at densities well below $10^{11}$ g cm$^{-3}$ but its strength $Q_\mathrm{sh}$,
as well as the depth at which it is deposited, appear to vary significantly from one source to another.
Typical values needed for $Q_\mathrm{sh}$ are of the order of 1--3 MeV, but in some cases (XTE J1701-462, \citealt{Page:2013aa}, and Swift J174805.3-244637, \citealt{Degenaar:2015aa}) it was not found to be required, while in the case of MAXI J0556-332 \citep{Deibel:2015aa} it could be well above 10 MeV.

%%%%%%%%%%%%%%%%%%%%%%%%%%%%%%%%%%%%%%%%%%%%%%%%%%%%%%%%%%%%%%%%%%%%%%%%%%%%%%%%
\subsection{MAXI J0556--332}

In the present work we focus  on the thermal evolution of the neutron star in MAXI J0556--322, an X-ray transient that was discovered in early 2011 \citep{Matsumura:2011tj}
when it had started a major outburst.
It was in outburst for 16 months before it returned to quiescence in May of 2012. Although no thermonuclear bursts or pulsations were detected, the behavior of the source in an X-ray hardness-intensity diagram strongly suggested that the accreting compact object in \MAXI\ is a neutron star: the tracks traced out at the highest fluxes showed a strong resemblance to those of the most luminous neutron star low-mass X-ray binaries, the Z sources. Based on a flux comparison with other Z sources, \citet{homan2011} suggested that the source is a very distant halo source. 

\MAXI\ has shown three smaller outbursts since its discovery outburst, in late 2012, 2016, and 2020. 
\citet{homan2014} studied the cooling of the neutron star after the first two outbursts. 
They found that during the first few years in quiescence after the 2011/2012 outburst, the neutron star in \MAXI\ was exceptionally hot compared to the other cooling neutron stars that have been studied, as can be seen in Figure \ref{fig:Cool_All}. The smaller second outburst that started in late 2012 and lasted $\sim$115 days did not produce detectable deviations from the cooling trend seen after the first outburst. \citet{Deibel:2015aa} showed that to produce the extremely high temperatures observed after the first outburst, a high amount of shallow heating was required ($\sim$6--16 MeV per accreted nucleon). They further concluded that the shallow heating mechanism did not operate during the second outburst. \citet{Parikh:2017uj} analyzed the neutron star cooling observed after the first three outbursts. Reheating of the crust was observed after the third outburst. It was concluded that the strength of the shallow heating in \MAXI\ varied from outburst to outburst. For two other sources the cooling has been studied after multiple outbursts as well. In MXB 1659--29 the strength of the shallow heating was found to be constant in the two outbursts after which cooling was studied. For Aql X-1, with numerous but short outbursts, the shallow heating was found to differ in both strength and depth between  outbursts \citep{Degenaar:2019aa}. 
A definitive explanation for a varying shallow heating in \MAXI\ and Aql X-1 has not been provided yet.

Here we present a re-modelling of the crustal cooling data of \MAXI, including data taken after the end of the most recent outburst from 2020.
We argue that the high crustal temperatures caused by the 2011/2012 outbursts were not the result of anomalously strong shallow heating, but were instead caused by a gigantic thermonuclear explosion that occurred at some time during the last three weeks of that outburst. As this explosion was much more energetic than any other thermonuclear X-ray burst observed to date, even about a hundred times more powerful than a superburst, we find it appropriate to call it a ``hyperburst''.
We infer it must have been produced by unstable thermonuclear burning of neutron rich isotopes of oxygen or neon.

The paper is organized as follows. 
In \Sec{Sec:2020outburst} we present the analysis of the data taken during and soon after the 2020 outburst of \MAXI. 
In \Sec{Sec:Modeling} we briefly describe how we model the neutron star temperature evolution
and in \Sec{Sec:shallow} we compare two different scenarios of shallow heating.
In section \Sec{Sec:hburst} we propose the new scenario of a hyperburst as the cause of the hot crust of \MAXI\ when it exited the first 2011-2012 observed outburst and in 
\Sec{Sec:fuel} we try to identify the source of this explosion.
We discuss our results in \Sec{Sec:discussion} and conclude in \Sec{Sec:conclusions}.

%-----------------------------------------------------------------------------------------------
\begin{figure}[t]
    \centering
    \includegraphics[width=8.5cm]{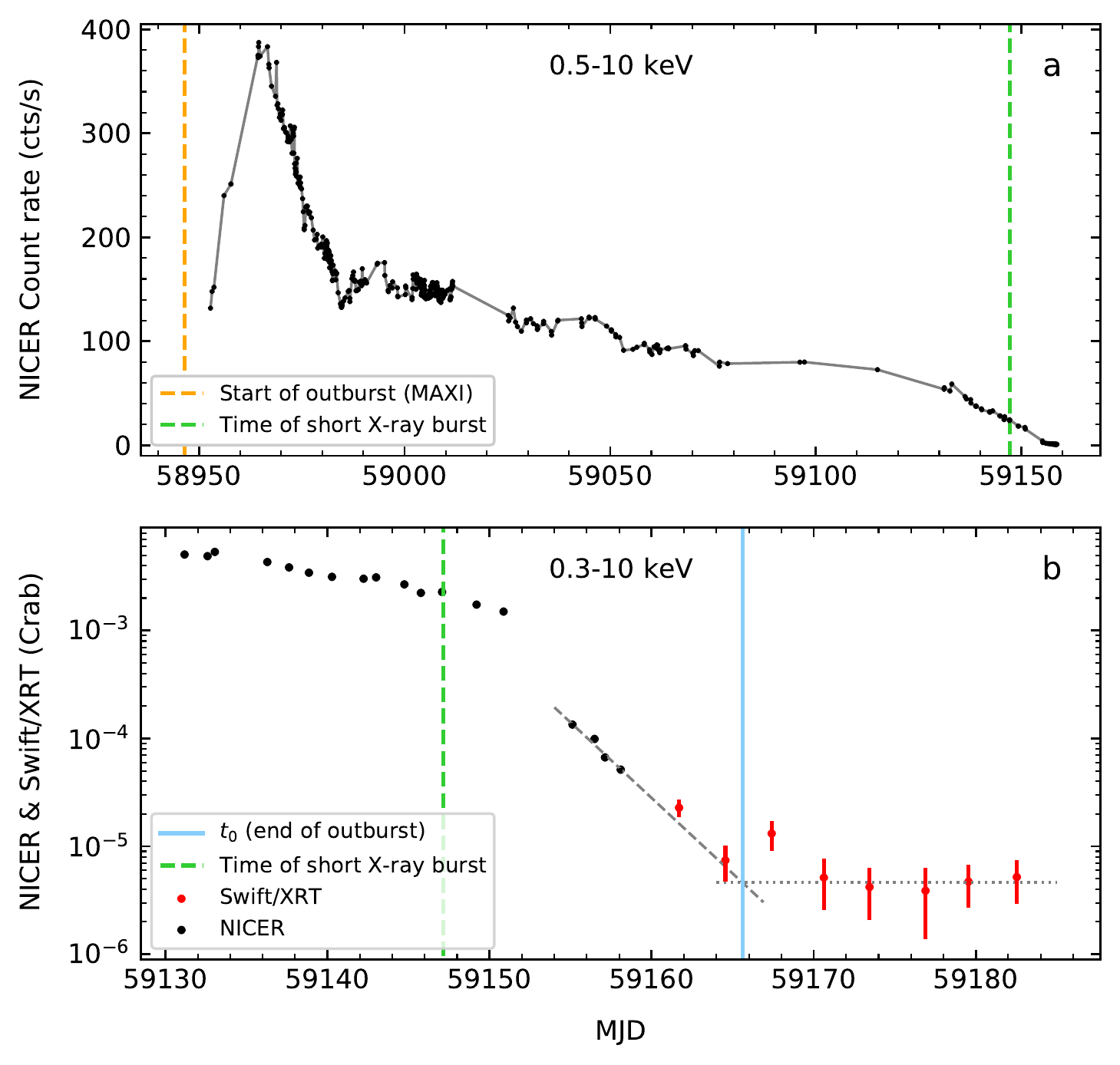}
    \caption{(a): {\it NICER} 0.5--10 keV light curve of the 2020 outburst of \MAXI. Each data point represents a single GTI. Background was not subtracted. The start of the outburst (as observed with MAXI) and the occurrence of a small X-ray burst are marked with dashed orange and green lines, respectively. (b) {\it NICER} (black) and {\it Swift} (red) coverage of the decay into quiescence. Each data point represents the background subtracted 0.3--10 keV count rate in Crab units. The blue line shows the estimate of the end of the outburst (MJD 59165.6), obtained by finding the intersection between an exponential fit to the last four {\it NICER} and first two {\it Swift} data points, and a constant fit to the last five {\it Swift} data points.}
    \label{fig:outburst}
\end{figure}
%-----------------------------------------------------------------------------------------------

%%%%%%%%%%%%%%%%%%%%%%%%%%%%%%%%%%%%%%%%%%%%%%%%%%%%%%%%%%%%%%%%%%%%%%%%%%%%%%%%
\section{The 2020 Accretion Outburst} 
\label{Sec:2020outburst}
%%%%%%%%%%%%%%%%%%%%%%%%%%%%%%%%%%%%%%%%%%%%%%%%%%%%%%%%%%%%%%%%%%%%%%%%%%%%%%%%

%%%%%%%%%%%%%%%%%%%%%%%%%%%%%%%%%%%%%%%%%%%%%%%%%%%%%%%%%%%%%%%%%%%%%%%%%%%%%%%%
\subsection{NICER observations}\label{sec:nicer}

The 2020 outburst of \MAXI\ was monitored extensively with the X-ray Timing Instrument (XTI) on board the Neutron Star Interior Composition Explorer \citep[{\it NICER;}][]{gendreau2016}. The XTI provides coverage in the 0.2--12 keV band and consists of 56 non-imaging X-ray concentrators that are coupled to Focal Plane Modules (FPMs), each containing a silicon drift detector. At the time of the observations 52 of the 56 FPMs were functional. A total of 107 ObsIDs are available for the 2020 outburst, each with one or more good-time intervals (GTI). Data were reprocessed with the {\tt nicerl2} tool that is part of HEASOFT v6.29. Default filtering criteria\footnote{\url{https://heasarc.gsfc.nasa.gov/docs/nicer/analysis_threads/nicerl2/}} were used. We additionally used the count rate in the 13--15 keV band, where no source contribution is expected, to filter out episodes of increased background (13–-15 keV count rates $>$1.0 counts s$^{-1}$ per 52 FPMs). After filtering, a total exposure of $\sim$332 ks remained. 

A full 0.5--10 keV outburst light curve was made with one data point per GTI. The GTIs varied in length from 16 s to $\sim$2.1 ks. The outburst light curve is shown in Figure \ref{fig:outburst}a. {\it NICER} observations started six days after the first MAXI/GSC detection of source activity \citep[][see orange dashed line in Figure \ref{fig:outburst}]{negoro2020} and caught the source during a fast rise towards the peak of the outburst. After the peak of the outburst the {\it NICER} count rate dropped rapidly for $\sim$20 days, after which the count rate decreased more slowly until the end of the outburst, about half a year later. 

\begin{figure}[!t]
    \centering
    \includegraphics[width=8.5cm]{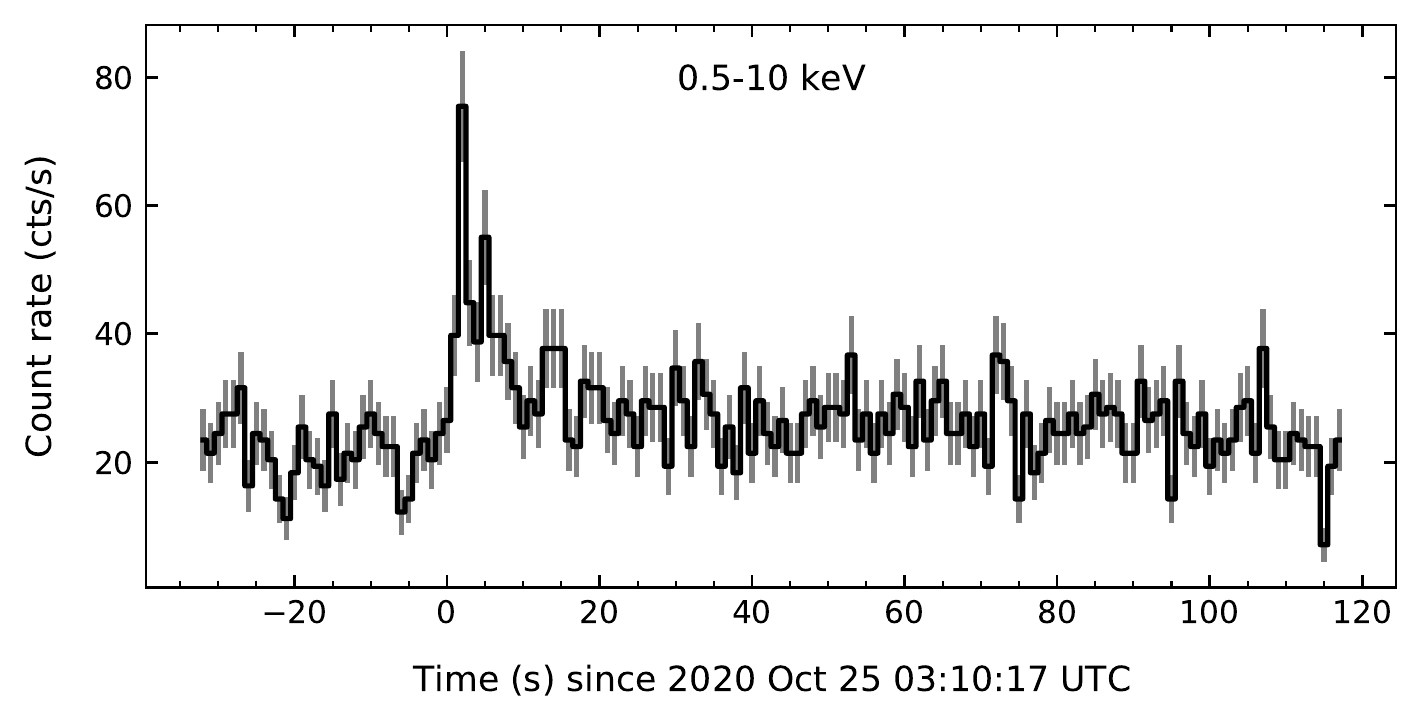}
    \caption{A 0.5--10 keV light curve of \MAXI\ (ObsID 3201400198) showing a short X-ray bursts. The time resolution is 1 s.}
    \label{fig:burst}
\end{figure}

We also produced 0.5-10 keV light curves for each ObsID with a time resolution of 1 s to search for possible X-ray bursts. One candidate X-ray burst was found in ObsID 3201400198 (MJD 59147, 2020 October 25). A segment of the light curve of the observation is shown in Figure \ref{fig:burst}. The burst is very short ($<$10 s from start to end) and peaks at a count rate a factor $\sim$3 higher than the persistent emission. Fits to the spectrum of the persistent emission slightly favor a soft spectral state at the time of the burst. Fitting a ``double-thermal" model \citep{lin2007} with $n_{\rm H}$ fixed to 3.2$\times10^{20}$ cm$^{-2}$ \citep{Parikh:2017aa} gives black-body and disk-black-body temperatures of 1.45(1) keV and 0.42(1), respectively, with a reduced $\chi^2$ of 1.00 (a typical hard state model, black-body with power-law, resulted in a slightly worse reduced $\chi^2$ of 1.07). The unabsorbed 0.5--10 keV flux was 5.8$\times10^{-11}$ erg\,cm$^{-2}$\,s$^{-1}$, corresponding to a luminosity of 1.3$\times10^{37}$ erg\,s$^{-1}$ for a distance of 43.6 kpc \citep{Parikh:2017aa}. Careful inspection of the data suggests that the burst is not the result of a short, sudden increase in the background. The burst is also present in various subsets of detectors, indicating that the burst was not instrumental. The count rates during the burst are too low to perform a study of the spectral evolution, but the burst was more pronounced at higher energies ($>$2 keV), which is consistent with a thermonuclear (type I) X-ray burst. We used WebPIMMS\footnote{\url{https://heasarc.gsfc.nasa.gov/cgi-bin/Tools/w3pimms/w3pimms.pl}} to convert the peak count rate (75 counts\,s$^{-1}$) into a 0.01--100 keV luminosity (assuming a black-body  temperature of 2.5 keV, a distance of 43.6 kpc, and an $n_{\rm H}$ of 3.2$\times10^{20}$ cm$^{-2}$):  $\sim3.8\times10^{38}
$ erg\,s$^{-1}$, which is very close to the empirical Eddington limit of neutron stars \citep{kuulkers2003}. This would constitute the first detection of a type I X-ray burst from \MAXI\ and would add further proof that the compact object is a neutron star.

Background subtracted spectra were produced for each ObsID using the {\tt nicerbackgen3C50} tool \citep{remillard2021}, which uses several empirical parameters to construct background spectra from a library of observations of {\tt NICER} background fields. FPMs 14 and 34 were excluded, since they tend to be affected by noise more often than the other detectors. We note that background spectra could only be extracted for 76 of the 107 ObsIDs; for the remaining ones the empirical parameters fell outside the range covered by the library background observations. From the resulting spectra a background-subtracted 0.3--10 keV light curve was produced. In \S \ref{sec:swift} this light curve will be used in conjunction with {\it Swift} data to estimate the end of the outburst and the start of quiescence.

%-----------------------------------------------------------------------------------------------
%-----------------------------------------------------------------------------------------------

%%%%%%%%%%%%%%%%%%%%%%%%%%%%%%%%%%%%%%%%%%%%%%%%%%%%%%%%%%%%%%%%%%%%%%%%%%%%%%%%
\subsection{Swift observations}\label{sec:swift}

The {\it Neil Gehrels Swift Observatory} X-Ray Telescope \citep[{\it Swift}-XRT;][]{burrows2005} observed \MAXI\ eight times near the end of the 2020 outburst. A 0.3--10 keV light curve with one data point per ObsID was produced by the online {\it Swift}-XRT products generator\footnote{\url{https://www.swift.ac.uk/user_objects/}} \citep{evans2007,evans2009}. Background was not subtracted, since the corrections were expected to be only minor ($\sim$5\%). The {\it Swift} light curve (red data points) is shown together with the {\it NICER} background-subtracted data (black) in Figure \ref{fig:outburst}b. Both the {\it Swift} and {\it NICER} data were normalized to Crab units. 

The {\it NICER} data in Figure \ref{fig:outburst}b cover a switch from a slow to a rapid decay, around MJD 59153. The {\it Swift} data covered the end of the decay and the start of quiescence, as indicated by the levelling off of the count rates. To estimate the start of quiescence ($t_0$) we determined the intersection of the decay (modeled with an exponential) and the quiescent level (modeled with a constant).  A similar method was previously employed for other sources \citep[see][]{Fridriksson:2010vx,Waterhouse:2016ab,Parikh:2017aa,Parikh:2019aa}. The exponential fit was made to the last four {\it NICER}  and the first two {\it Swift} data points. The constant fit was made to the last five {\it Swift} data points. The third {\it Swift} data point was excluded since the higher count rate indicated a possible short episode of enhanced accretion, which was also observed several times soon after the end of the 2011/2012 outburst \citep{homan2014}. From the fits we find $t_0$=MJD 59165.6. This time is marked with a blue line in Figure \ref{fig:outburst}b, along with the exponential (dashed line) and constant fits (dotted line).

Again using the online {\it Swift}-XRT products generator we extracted a single background-subtracted quiescent spectrum for the last five {\it Swift} observations. The spectrum had an exposure time of $\sim$12.6 ks, contained $\sim$30 counts, and was fit with XSPEC v12.12 \citep{ar1996} using C-statistics. We used a similar absorbed neutron star atmosphere model as in \citet{homan2014} and \citet{Parikh:2017uj}: {\tt constant $\times$ tbabs $\times$ nsa}. Note that these authors employed {\tt nsa} instead of the frequently used neutron star atmosphere model {\tt nsatmos} \citep{heinke2006}, since the latter model does not allow temperatures above $\mathrm{log}_{10}(kT)=6.5$. {\tt constant} is set to a value of 0.92 to account for cross-calibration with other X-ray detectors.  Although only a {\it Swift} spectrum is analyzed in our work, the $n_{\rm H}$ and distance used in our fits are based  on the {\it Chandra} and {\it XMM-Newton} spectra fitted by \citet{Parikh:2017uj}; not using the cross-calibration constant would result in a temperature that is too low. The absorption component {\tt tbabs} is an updated version of the {\tt tbnew\_feo} component used by \citet{homan2014}, but it yields the same column densities. We set the abundances to {\tt WILM} and cross sections to {\tt VERN}, and fixed the $n_{\rm H}$ to the value obtained by \citet{Parikh:2017uj}: 3.2$\times10^{20}$ cm$^{-2}$. For the neutron star atmosphere component {\tt nsa} we fixed the neutron star mass ($M$) and radius\footnote{In this paper by radius we always mean the coordinate radius in Schwarzschild coordinates, i.e., {\it not} the ``radius at infinity''.} ($R$)
to 1.4 $M_\odot$ and 10 km, respectively, and for the normalization we used a distance of 43.6 kpc.  The only free parameter in the fit was the temperature of the {\tt nsa} component. When converted to the effective temperature measured at infinity\footnote{$T_\mathrm{eff}^{\infty}=T_\mathrm{eff}/(1+z)$, where $(1+z) \equiv (1- R_{S}/R)^{-1/2} $ is the gravitational redshift factor, with $R_S=2GM/c^2$ being the Schwarzschild radius.} we find a value of 166$\pm$8 eV. This data point (for which we use the midpoint of the five observations as date: MJD 59176.6) was added to the cooling data obtained by \citet{Parikh:2017uj} from prior {\it Swift}, {\it Chandra}, and {\it XMM-Newton} observations. 

%%%%%%%%%%%%%%%%%%%%%%%%%%%%%%%%%%%%%%%%%%%%%%%%%%%%%%%%%%%%%%%%%%%%%%%%%%%%%%%%
\section{Modeling the Evolution of the \MAXI\ Neutron Star} 
\label{Sec:Modeling}
%%%%%%%%%%%%%%%%%%%%%%%%%%%%%%%%%%%%%%%%%%%%%%%%%%%%%%%%%%%%%%%%%%%%%%%%%%%%%%%%

We build on our previous experience in modeling accretion heated neutron stars using an updated version of the code \texttt{NSCool} \citep{Page:2016aa}, 
which solves the general relativistic equations of stellar structure and thermal evolution.
The physics we employ is standard and briefly described in \App{Sec:physics}.

To model the mass accretion rate during the four observed outbursts we apply a similar method as previously described in \cite{ootes2016,Ootes:2018aa} and \cite{Parikh:2017uj}. However, instead of using a mix of {\it Swift}, {\it NICER} and {\it MAXI} data to estimate the evolution of the bolometric flux, we opt to only use the {\it MAXI} data, since they provide a single consistent set that covers all four outbursts and requires just one count-rate-to-flux conversion factor. For the 2--20 keV MAXI count rate to bolometric flux conversion factor we used the value from \citet{Parikh:2017uj}: 2.353$\times10^{-8}$ erg\,cm$^{-2}$\,count$^{-1}$.

From the bolometric flux $F_\mathrm{bol}$ we calculate the daily average mass accretion rate using
\be
\dot M = \frac{F_\mathrm{bol}4\pi D^2}{\eta c^2}
\label{Eq:Mdot}
\ee
where $D$ is the source distance, fixed here at $D = 43.6$ kpc, $c$ the speed of light and $\eta \approx 0.2$ the accretion efficiency factor \citep{Shapiro:1986wz}.
Heating was assumed to only take place during the outburst intervals, which were fixed by hand and are shown as gray shaded areas in Figure \ref{fig:Bigbadaboom}. 
The temporal evolution of the heating luminosities, $H_\mathrm{dc}(t)$ for the deep crustal heating and $H_\mathrm{sh}(t)$ for the shallow heating,
follow the instantaneous mass accretion rate as
\be
\left( \begin{array}{cc} H_\mathrm{dc}(t) \\ H_\mathrm{sh}(t) \end{array} \right) =
\left( \begin{array}{cc} Q_\mathrm{dc} \\ Q_\mathrm{sh} \end{array} \right) \times \frac{\dot M(t)}{m_u}
\label{Eq:Heating}
\ee
with their respective strengths $Q_\mathrm{dc}$ and $Q_\mathrm{sh}$.
We follow \citet{Haensel:2008aa} and use $Q_\mathrm{dc} = 1.93$ MeV that is distributed in density at the locations of each reaction (see figure 3 in 
\citealt{Haensel:2008aa}) for the model starting with $^{56}$Fe at low densities.
For the shallow heating we assume $Q_\mathrm{sh}$ is distributed uniformly, per unit volume, within a shell covering a density range from $\rho_\mathrm{sh}$
up to $5 \rho_\mathrm{sh}$. 
Both $Q_\mathrm{sh}$ and $\rho_\mathrm{sh}$ are considered, in a purely phenomenological manner, as free parameters that are adjusted to fit the data.

In our model of a thermonuclear hyperburst we assume an energy $X_\mathrm{hb} \times 10^{18}$ erg g$^{-1}$ is deposited almost instantaneously
at time $t_\mathrm{hb}$ in a shell going from our outer boundary layer (at $\rho = \rho_b = 10^8$ g cm$^{-3}$) up to a density $\rho_\mathrm{hb}$.
We envision the layer at density $\rho_\mathrm{hb}$ as a ``critical layer'' and the latter as a ``critical density'', being the point where the hyperburst is triggered.
The temperature of this critical layer at the moment when the hyperburst is triggered,  $T_\mathrm{hb}$, is not fixed in any way but will be an output of our simulations
and will be referred to as a ``critical temperature''.
$X_\mathrm{hb}$, $t_\mathrm{hb}$, and $\rho_\mathrm{hb}$, are also treated phenomenologically as free parameters to fit the data.
An energy of $10^{18}$ erg g$^{-1}$ corresponds to about 1 MeV per nucleon and is a typical energy released by fusion of C, O or Ne into iron-peak nuclei
(see \Tab{tb:Q})
so that $X_\mathrm{hb}$ roughly represents the mass fraction of the exploding nuclear species.

In the previous study of the evolution of \MAXI\ in \citet{Parikh:2017uj} a $\chi^2$ minimisation technique was used to obtain the best fit to the data.
Here, however, we apply the more robust method, for models with many parameters, of Markov Chain Monte Carlo (MCMC) using our recently developed
MCMC driver \texttt{MXMX} (see \citealt{Lin:2018aa,Ootes:2019aa,Degenaar:2021us}).
The parameters we vary are the mass $M$ and radius $R$ of the star, 
which determine the crust thickness that is a dominant factor in controlling, on one hand, 
the time scale for heat transport and, on the other hand, the amount of matter present and thus the amount of energy needed to heat it\footnote{Notice that our spectral fits to deduce $T_\mathrm{eff}$ were performed assuming $M=1.4 M_\odot$ and $R=10$ km: for full consistency with our MCMC runs these should be performed for a range of $M$ and $R$.
%as, e.g., was done in \citealt{Degenaar:2021us}. 
These, we expect, would result in shifts of $T_\mathrm{eff}$ of the order of 10\% and should not have major impact on our conclusions.},
the initial red-shifted core temperature $\widetilde{T}_0$, 
the depth of the light element layer in the stellar envelope, $y_\mathrm{L}$, which affects the observed effective temperature $T_\mathrm{eff}$,
the impurity parameter $Q_\mathrm{imp}$ that can strongly reduce the thermal conductivity in the solid crust,
the strength and depth of the shallow heating, $Q_\mathrm{sh}$ and $\rho_\mathrm{sh}$,
and, when considered, the properties of the hyperburst, $X_\mathrm{hb}$, $t_\mathrm{hb}$, and $\rho_\mathrm{hb}$.
In \App{Sec:MCMC} we describe the various settings we apply in the various scenarios presented in the following sections.

%%%%%%%%%%%%%%%%%%%%%%%%%%%%%%%%%%%%%%%%%%%%%%%%%%%%%%%%%%%%%%%%%%%%%%%%%%%%%%%%
\section{Shallow Heating in the Four Observed Accretion Outbursts} 
\label{Sec:shallow}
%%%%%%%%%%%%%%%%%%%%%%%%%%%%%%%%%%%%%%%%%%%%%%%%%%%%%%%%%%%%%%%%%%%%%%%%%%%%%%%%

%------------------------------------------------------------------------------------------------
\begin{figure}[t]
	\begin{center}
	\includegraphics[width=0.99\columnwidth]{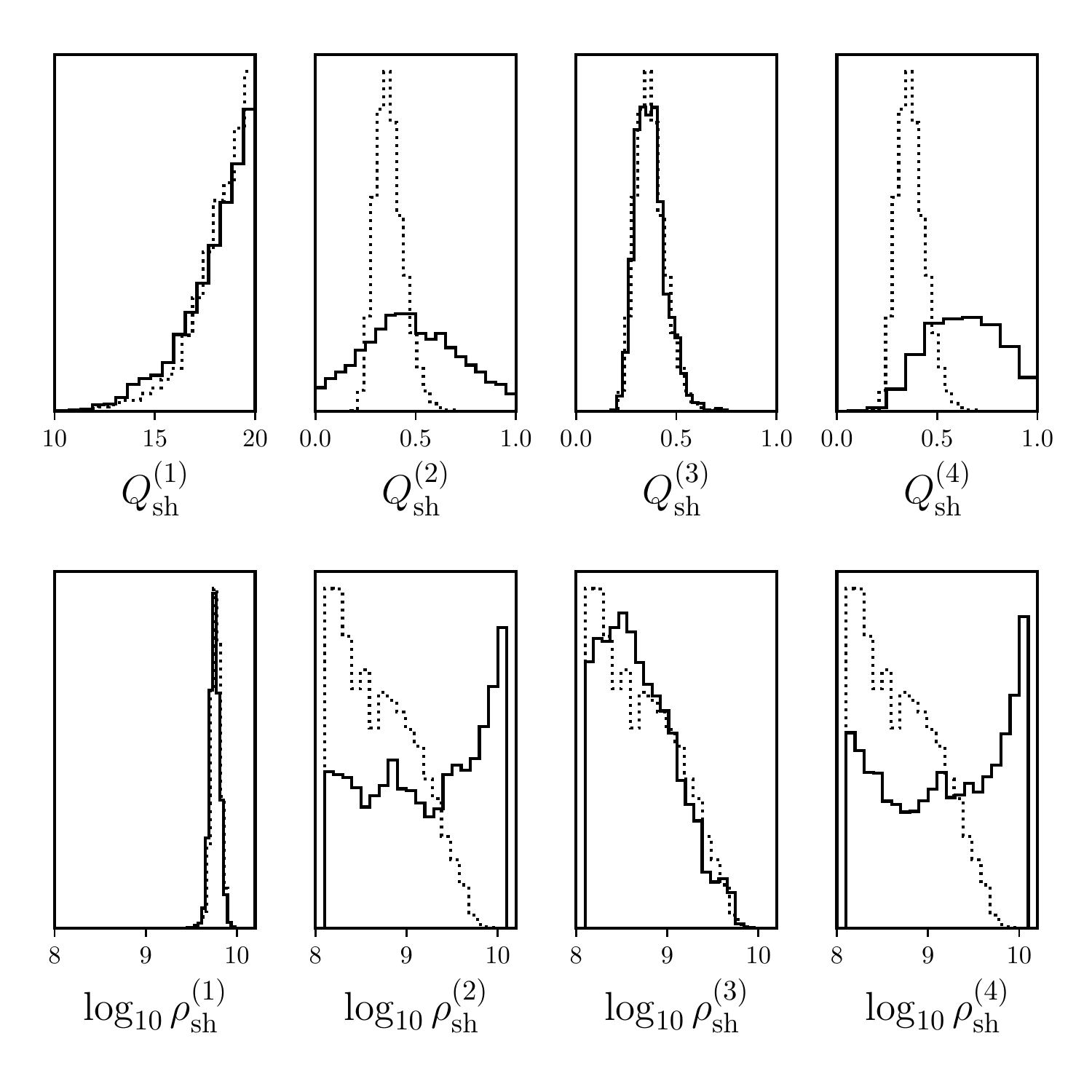}
	\end{center}
	\caption{Histograms of the posterior distributions of the strength, $Q_\mathrm{sh}^{(i)}$ in MeV, 
	and lower density, $\rho_\mathrm{sh}^{(i)}$ in g cm$^{-3}$, 
	of shallow heating during outburst $i$ ($i=1, \dots, 4$) for Scenario ``A'', continuous lines, and ``B'', dotted lines.
	(Vertical scales are linear and adjusted such that the two histograms cover the same area.)
	}
	\label{fig:Qsh_AB}
\end{figure}
%------------------------------------------------------------------------------------------------

%------------------------------------------------------------------------------------------------
\begin{figure}[t]
	\begin{center}
	\includegraphics[width=0.99\columnwidth]{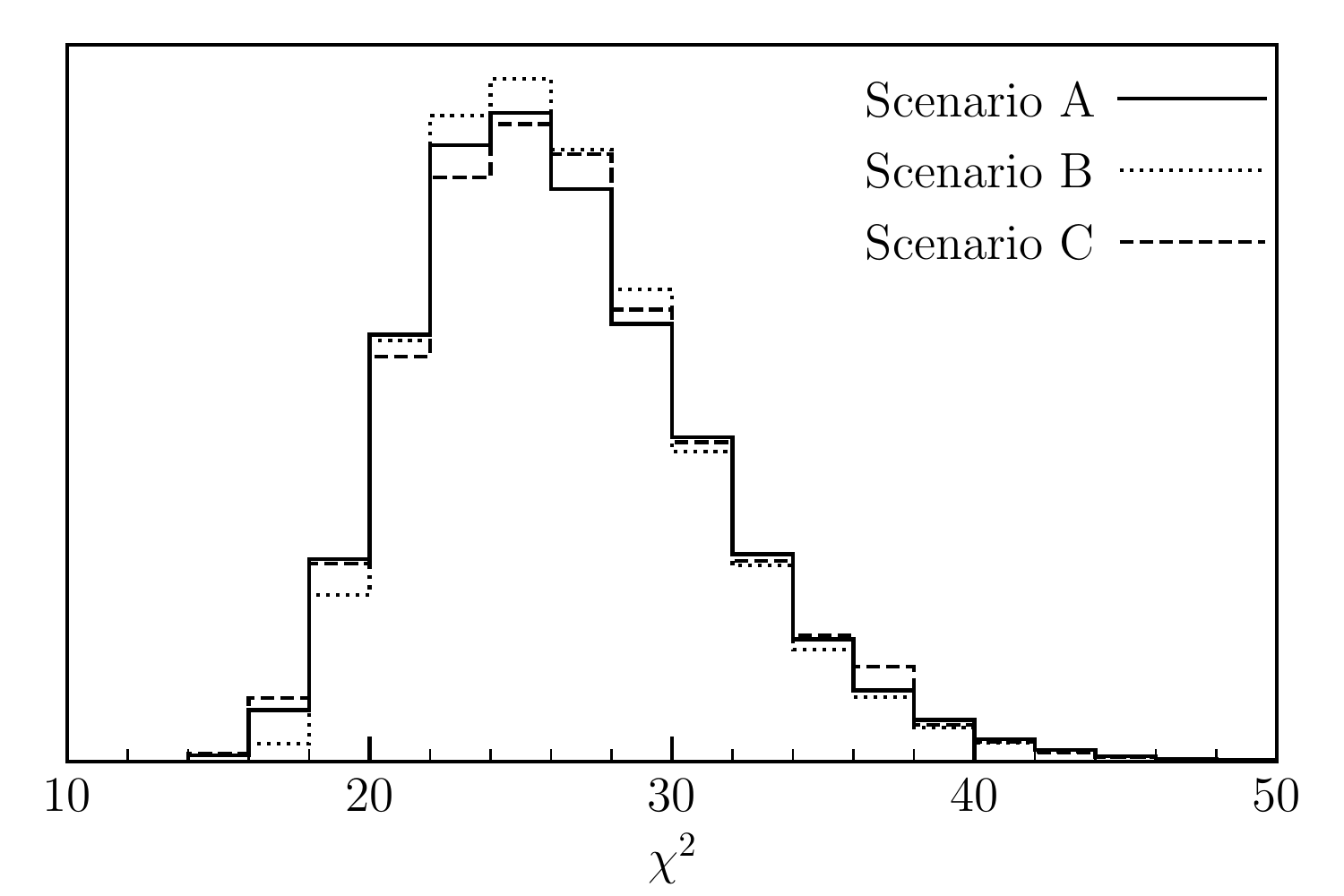}
	\end{center}
	\caption{Histograms comparing the $\chi^2$ distributions of our 3 scenarios ``A'', ``B'', and ``C''.
	We have about 2 millions samples in scenarios ``A'' and ``B'' and 4 millions in ``C''.
	}
	\label{fig:chi2_ABC}
\end{figure}
%-----------------------------------------------------------------------------------------------

To extend the work of \citet{Parikh:2017uj} and incorporate the new information from the 2020 accretion outburst we set up a first scenario, ``A'', 
with 20 parameters:
$M$, $R$, $T_0$, 5 zones with different $Q_\mathrm{imp}^{(i)}$ ($i=1, \dots, 5$) and, for each one of the four observed accretion outbursts
$y_\mathrm{L}^{(j)}$, $Q_\mathrm{sh}^{(j)}$ and $\rho_\mathrm{sh}^{(j)}$ ($j=1, \dots, 4$).
We allow $Q_\mathrm{sh}^{(1)}$ to take values up to 20 MeV while we restrict $Q_\mathrm{sh}^{(2),(3),(4)}$ to at most 1 MeV in light of the results of \citet{Parikh:2017uj}.
In all cases $\mathrm{log}_{10} \rho_\mathrm{sh}^{(j)}$ can take any value between 8.2 and 10.2 (in g cm$^{-3}$).
Under these premises, we ran our MCMC (see \App{Sec:MCMC} for details) and generated about 2 millions samples.
We present in \Fig{fig:Qsh_AB}, continuous lines, the posterior distribution of the eight parameters $Q_\mathrm{sh}^{(j)}$ and $\rho_\mathrm{sh}^{(j)}$.
There is a clear dichotomy between outburst 1 on one hand and the next three ones on the other hand, already clearly identified by \citet{Deibel:2015aa}
and \citet{Parikh:2017uj}, with $Q_\mathrm{sh}^{(1)} \ge 10$ MeV while  $Q_\mathrm{sh}^{(2,3,4)} \le 1$ MeV.
Notice that the values of $Q_\mathrm{sh}^{(1)}$ show no upper limit because, as found by \citet{Deibel:2015aa}, when it is much above 10 MeV the outer crust
becomes so hot that most of the shallow heating energy is lost to neutrinos and increasing $Q_\mathrm{sh}^{(1)}$ has no further effect on the temperature profile
so that an arbitrary cut-off has to be introduced (we choose 20 MeV).
However, the distributions of both $Q_\mathrm{sh}^{(j)}$ and $\mathrm{log}_{10} \rho_\mathrm{sh}^{(j)}$ in outbursts 2, 3, and 4 point to the possibility
that shallow heating had the same properties in these three outbursts.
To evaluate this possibility we consider a second scenario, ``B'', in which $Q_\mathrm{sh}^{(j)}$ and $\rho_\mathrm{sh}^{(j)}$ have the same
values in these three outbursts, $Q_\mathrm{sh}^{(j)} = Q_\mathrm{sh}^{(234)}$ and $\rho_\mathrm{sh}^{(j)} = \rho_\mathrm{sh}^{(234)}$ for $j=2,3,4$. 
The resulting posterior distributions, from a second MCMC run (see \App{Sec:MCMC} for details) generating also about 2 millios samples,
are plotted in \Fig{fig:Qsh_AB} as dotted lines.
To compare these two scenarios we plot in \Fig{fig:chi2_ABC} the resulting $\chi^2$ distributions: 
it is clear that there is no really significant difference between them and considering that scenario ``B'' has 16 parameters versus 20
in scenario ``A'', it seems there is no gain in considering that the shallow heating had different properties in the last three outbursts.
(Scenario ``C'' is described in the next section.)

It is important to notice that in scenario ``B'' the distribution of $Q_\mathrm{sh}^{(234)}$ is essentially determined by the third outburst in which both
scenarios result in the same posterior distribution.
Hopefully further observations of the relaxation of \MAXI\ in the near future will constrain more strongly the properties of shallow heating in the fourth
outburst and strengthen our present conclusion that data from outbursts 2, 3, and 4, are compatible with having identical 
$Q_\mathrm{sh}$ and $\rho_\mathrm{sh}$.

%%%%%%%%%%%%%%%%%%%%%%%%%%%%%%%%%%%%%%%%%%%%%%%%%%%%%%%%%%%%%%%%%%%%%%%%%%%%%%%%
\section{A Gigantic Thermonuclear Explosion in the First Accretion Outburst}
\label{Sec:hburst}
%%%%%%%%%%%%%%%%%%%%%%%%%%%%%%%%%%%%%%%%%%%%%%%%%%%%%%%%%%%%%%%%%%%%%%%%%%%%%%%%

%------------------------------------------------------------------------------------------------
\begin{figure*}
	\begin{center}
	\includegraphics[width=0.85\textwidth]{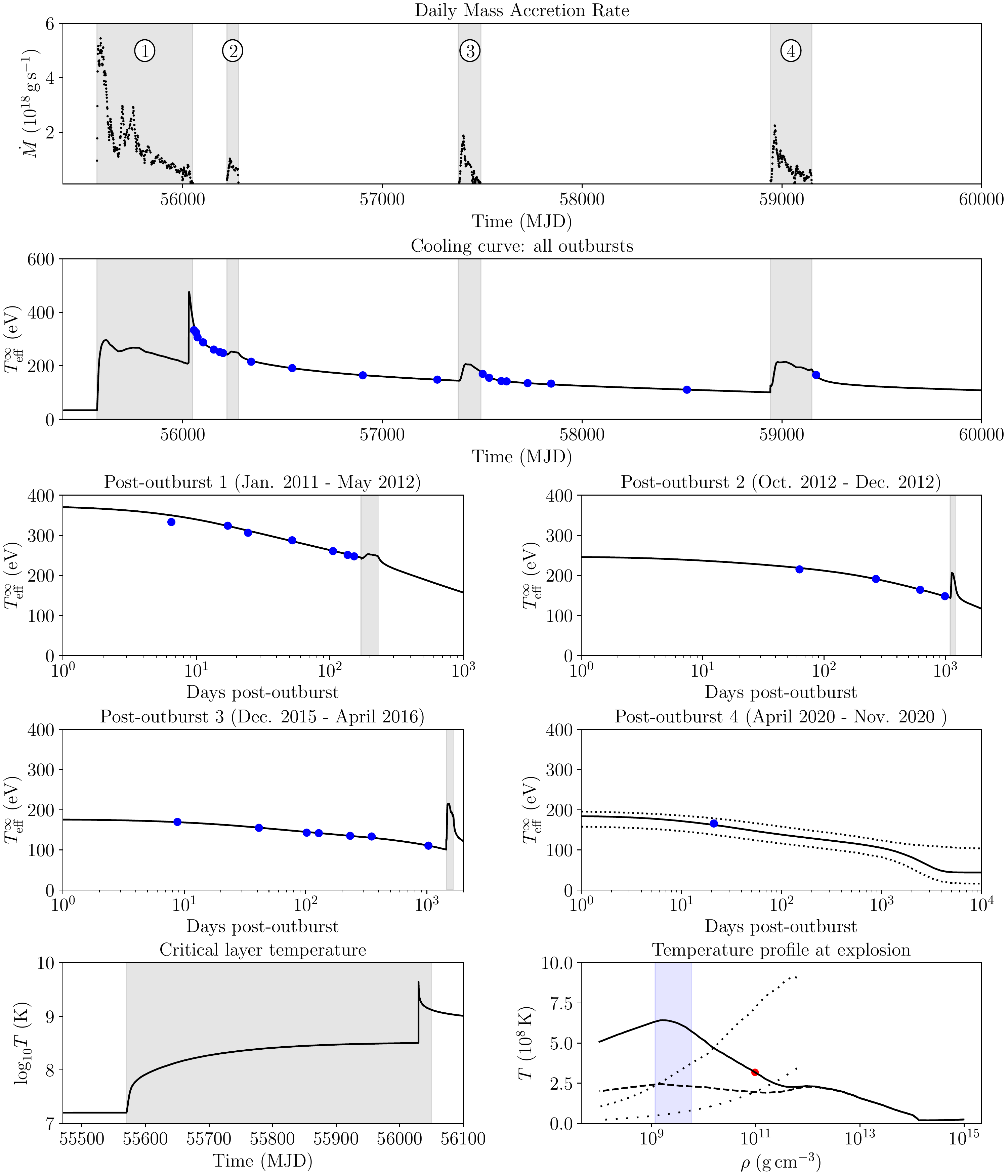}
	\end{center}
	\caption{
	Evolution of a 1.6 $M_odot$ model, with a radius of 11.2 km and an initial red-shifted core temperature $\widetilde{T}_0 = 1.2 \times 10^7$ K.
	Its shallow heating parameters are $Q_\mathrm{sh} = 0.390$ MeV distributed in the density range 
	$[1.15-5.7] \times 10^9$ g cm$^{-3}$.
	The two upper panels show the whole temporal evolution, spanning more than ten years and four accretion outbursts,
	of the daily mass accretion rate, $\dot M$, and red-shifted effective temperature, 
	$T_\mathrm{eff}^\infty$,
	while the four central panels show details of the relaxation 
	after each accretion outburst. (In parentheses are the initial and final dates of the outburst.)
	With 19 measurements of $T_\mathrm{eff}^\infty$ this model has a $\chi^2$ of 24.6.
	In all these panels the grey shaded areas delineate the periods of accretion.
	In the ``Post-outburst 4'' panel the two dotted lines show the 3$\sigma$ range of predictions from the whole set of our scenario ``C'' models.
	The lower left panel displays the evolution of the local temperature in the critical layer at density $\rho_\mathrm{hb} = 10^{11}$ g cm$^{-3}$
	during the first accretion outburst where the hyperburst occurred at time of 56030 MJD., i.e., three weeks before the end of the outburst.
	The lower right panel displays the local temperature profile in the whole star, continuous line, just before the time when the hyperburst will be triggered,
	where the red dot signals the position of the critical layer at $\rho_\mathrm{hb}$,
	while the dashed curve shows the temperature profile at the same time in the absence of shallow heating.
	The blue shaded density range shows the region where the shallow heating is assumed to take place.
	The dotted curves show the melting temperature of the main matter (upper dotted curve) and of 
	$^{28}$Ne 
	bubbles in pressure equilibrium with the main matter (lower dotted curve).
	}
	\label{fig:Bigbadaboom}
\end{figure*}
%------------------------------------------------------------------------------------------------

%------------------------------------------------------------------------------------------------
\begin{figure*}
	\begin{center}
	\includegraphics[width=0.95\textwidth]{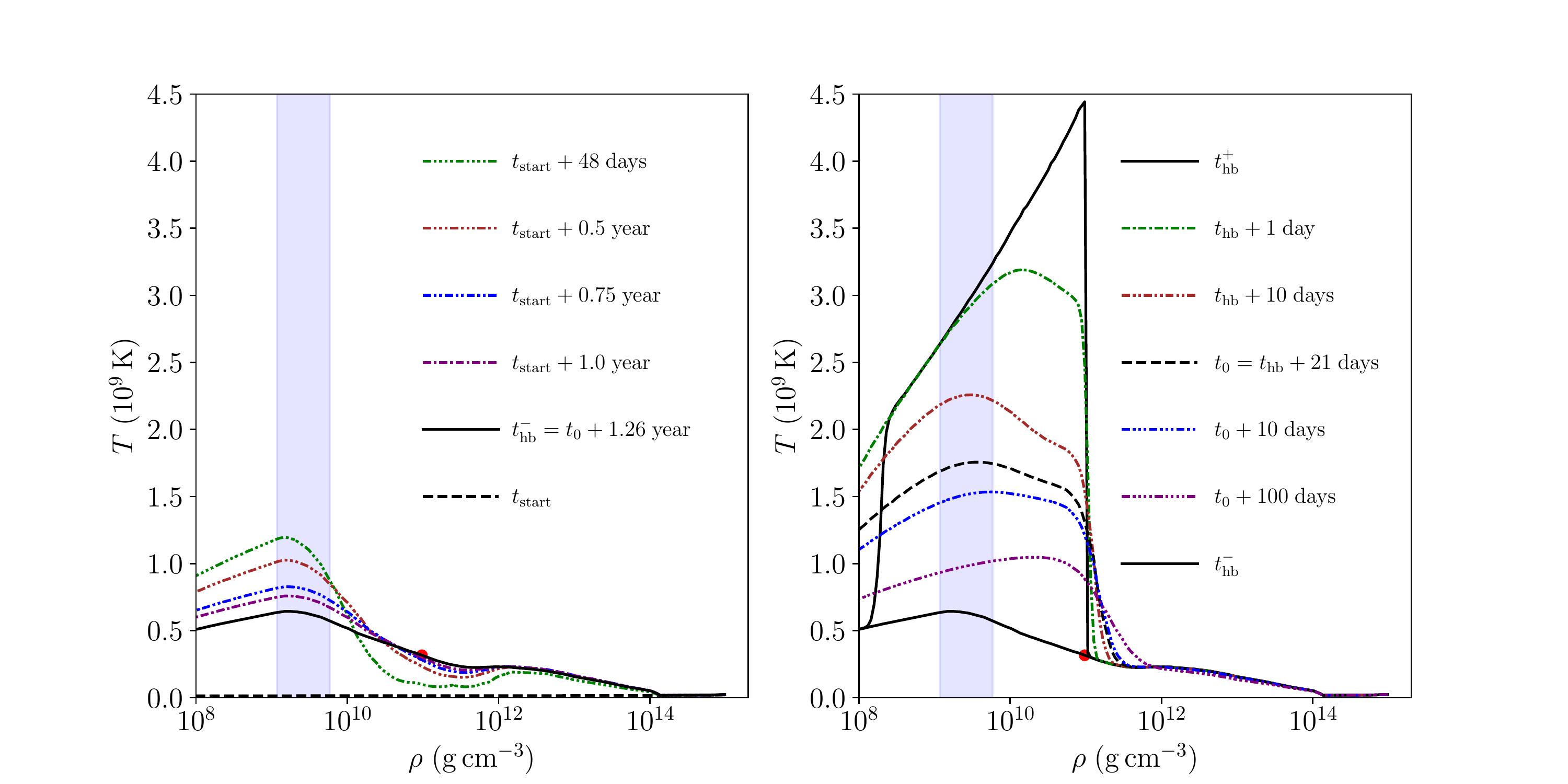}
	\end{center}
	\caption{Evolution of the neutron star internal temperature profile for the same model as in \Fig{fig:Bigbadaboom}.
	Left panel: profiles starting at the beginning of the first accretion outburst, $t_\mathrm{start}$, and five posterior times
	till just before the hyperburst explosion, $t_\mathrm{hb}^-$.
	Right panel: profiles starting just before, $t_\mathrm{hb}^-$, and after, $t_\mathrm{hb}^+$, the hyperburst explosion,
	with three posterior times till the end of the accretion outburst, $t_0$, and two later times during the quiescent phase.
	In both panels the blue shaded region marks the density range where the shallow heating is operating (during accretion).
	(Numerically, the hyperburst energy is smoothly injected during 200 seconds, the actual time interval from $t_\mathrm{hb}^-$
	to $t_\mathrm{hb}^+$. Moreover, no energy is injected close to the outer boundary at $10^8$ g cm$^{-3}$ but this anomaly
	is corrected by diffusion in less than a day as seen from the $t_\mathrm{hb} + \mathrm{1 \; day}$ profile.)
	}
	\label{fig:Bigbadaboom2}
\end{figure*}
%------------------------------------------------------------------------------------------------

%------------------------------------------------------------------------------------------------
\begin{figure*}
	\begin{center}
	\includegraphics[width=0.99\textwidth]{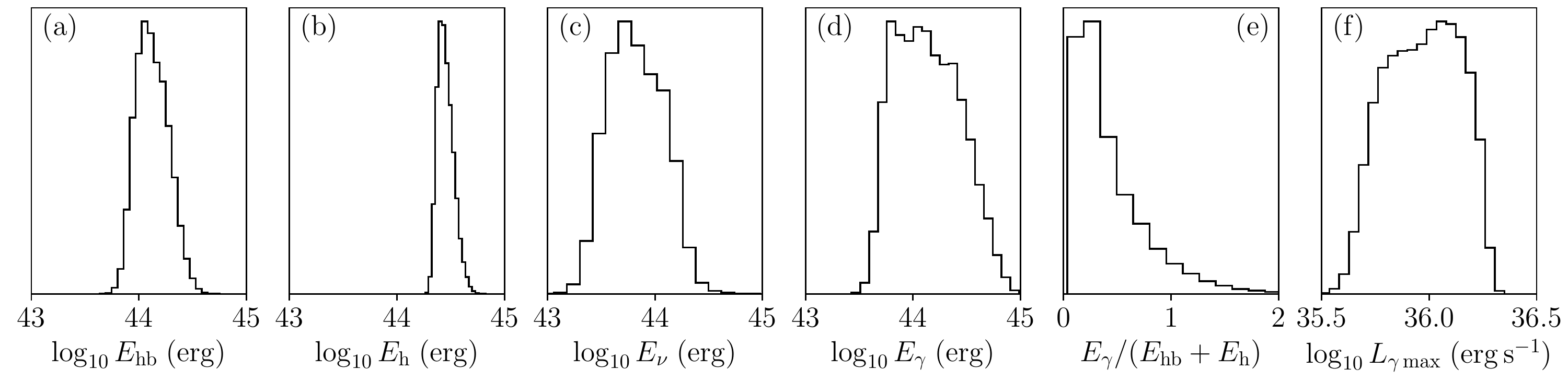}
	\end{center}
	\caption{Histograms of the distributions of some results from our scenario C MCMC:
	(a) total hyperburst energy, $E_\mathrm{hb}$, (b) total heating (shallow + deep crust), $E_\mathrm{h}$, during the four outbursts,
        total energy lost to (c) neutrinos, $E_\nu$, and (d) photons, $E_\gamma$, from the beginning of the first outburst till the end of the fourth one.
        Panel (e) shows the ratio of energy lost through photon emission to the total injected energy $E_\mathrm{hb}+E_\mathrm{h}$.
        The last panel, (f), shows the histogram of the maximum luminosity coming out of the stellar interior during the hyperburst.
	}
	\label{fig:ScenarioC_Energy}
\end{figure*}
%------------------------------------------------------------------------------------------------

Having shown that the properties of shallow heating were likely identical in outburst 2, 3, and 4 we now postulate that this was also the case during the first 
2011-2012 outburst and that the extremely high temperature of the neutron star at the end of that outburst is due to another phenomenon.
We explore here the possibility that this event may have been a gigantic thermonuclear explosion, triggered deep in the outer crust.
We, thus, present a third scenario, ``C'', with a single parametrization of the shallow heating, $Q_\mathrm{sh}$ and $\rho_\mathrm{sh}$, operating
in all four accretion outbursts and add a sudden heat injection with parameters $X_\mathrm{hb}$, $t_\mathrm{hb}$, and $\rho_\mathrm{hb}$,
as presented in \Sec{Sec:Modeling}, with $t_\mathrm{hb}$ adjustable to any time during the first outburst, whose duration was 1.31 year,
and $\rho_\mathrm{hb}$ allowed to take values between $10^8$ up to $10^{12}$ g cm$^{-3}$ and $X_\mathrm{hb}$ in the range $0 - 0.05$.

To explore scenario ``C'' we ran again our MCMC (see \App{Sec:MCMC} for details) and in \Fig{fig:chi2_ABC} the resulting distribution of the $\chi^2$ of this scenario is also exhibited.
There is no significant difference between the $\chi^2$ of the three scenarios, ``A'' and ``C'' have slightly larger distribution below 20 but this could simply
reflect the fact they have more parameters than ``B'' (20 and 17 versus 16, respectively).
This simple comparison indicates that
scenario C appears statistically at least as good as the first two.
  
In \Fig{fig:Bigbadaboom} we illustrate in detail one of the good models found in this manner.
The two upper panels display the overall evolution of both the daily mass accretion rate $\dot{M}$ and the star's red-shifted effective temperature $T_\mathrm{eff}^\infty$
while the four central ones exhibit the excellent fit to the 19 data points that we obtain for this more than a decade long evolution.
In the lower left panel of this figure one sees the constant rise in the temperature at the bottom of the exploding layer, at density $\rho_\mathrm{hb}$,
which could be consistent with this temperature reaching a critical value at which some nuclear burning becomes unstable resulting in a thermonuclear runaway.
We attempt to identify the fuel of the explosion in the next section.
In the lower right panel we present the temperature profile in the whole star at the time just before the explosion,
compared with a model where shallow heating was not implemented,
showing that the occurrence of shallow heating was crucial in setting this profile in the outer crust.
In \Fig{fig:Bigbadaboom2} we detail the evolution of the internal temperature, before (left panel) and after (right panel)
the hyperburst starting from the beginning of the first accretion outburst, at time $t_\mathrm{start}$. 
The profile at 48 days corresponds to the highest temperature in the region where the shallow heating is operating and is occurring
when the mass accretion rate is at its maximum.
In the subsequent profiles one notices the decreasing temperature of this low density region due to both the decrease of the mass accretion rate and the flow of heat inward, the latter resulting in a rising temperature in the
density range $10^{10}$ to $10^{12}$ g cm$^{-3}$: this explains the constant temperature rise at the point where
the explosion will be triggered, at $\rho_\mathrm{hb} = 10^{11}$ g cm$^{-3}$, as seen in the lower left panel of \Fig{fig:Bigbadaboom}.
Notice that in the low density layers, due to their low heat capacity and resulting short thermal timescales, the temperature
follows the variations of the mass accretion and heating rates. At higher densities the heating has a more cumulative 
effect and the temperature keeps rising while the mass accretion rate decreases 
(a similar evolution can be clearly seen in the ``Supporting Information'' movie of \citealt{ootes2016}).
The right panel of \Fig{fig:Bigbadaboom2} illustrates the relaxation of the temperature after the heat injection from the 
hyperburst: due to a fixed energy injection per gram the initial profile peaks at the highest density $\rho_\mathrm{hb}$.
The initial rapid drop in temperature is due to neutrino losses \citep{Deibel:2015aa} and
during the relaxation under accretion, i.e., at $t$ before $t_0$, the peak moves back toward the 
shallow heating region.
After the end of the accretion the absence of heating and the heat flow toward the surface result in the maximum temperature peak moving toward higher densities.
As a result of the inward heat flow, a large part of the energy released by the explosion is being stored at higher densities 
and will result in a very long cooling time: 
one sees in the ``Cooling curve'' panel of \Fig{fig:Bigbadaboom} that even after the perturbation from the third outburst
the cooling trajectory is a continuation of the trajectory from the first outburst   \citep{Parikh:2017uj}.

The global energetics of scenario ``C'' resulting from our MCMC are presented in \Fig{fig:ScenarioC_Energy}.
We find that a typical hyperburst energy $E_\mathrm{hb}$ is of the order of $10^{44}$ ergs, which is about 2 orders of magnitude larger than the one of a superburst \citep{Cumming:2006uv} and is comparable to the energy output of a magnetar Giant Flare \citep{Kaspi:2017ts}.
However, the total energy injected into the star from both the shallow and the deep crustal heating, $E_\mathrm{h}$, is larger than $E_\mathrm{hb}$ by a factor
of a few and the energy lost through photon emission, $E_\gamma$, is of comparable magnitude:
in most cases, however, $E_\gamma$ is somewhat smaller than the total heating energy $E_\mathrm{hb} + E_\mathrm{h}$.
The energy not lost to photons is stored into the core and neutrino losses will contribute to the global energy balance in the long term \citep{Brown:1998aa,Colpi:2001vn}, but a detailed study of this issue is beyond the scope of the present work.
Finally, the last panel of \Fig{fig:ScenarioC_Energy} shows the distribution of the peak luminosity $L_{\gamma \, \mathrm{max}}$ 
coming from the stellar interior during the hyperburst that is radiated as photons from the surface:
unfortunately this luminosity $L_{\gamma \, \mathrm{max}}$ is always significantly smaller than the X-ray luminosity inferred from the observed flux, 
$\sim 10^{38}$ erg s$^{-1}$ for a distance of 43.6 kpcs as we assume here, and thus unlikely to be noticeable in the data.
However, since we do not model the explosion at low densities, our outer boundary being at density $\rho_b = 10^8$ g cm$^{-3}$,
we cannot exclude that a hyperburst may trigger a lower density X-ray burst, in a manner similar to the superburst precursors
(see, e.g., \citealt{Galloway:2021vr}), which could be detectable.

\begin{figure}
	\begin{center}
	\includegraphics[width=0.99\columnwidth]{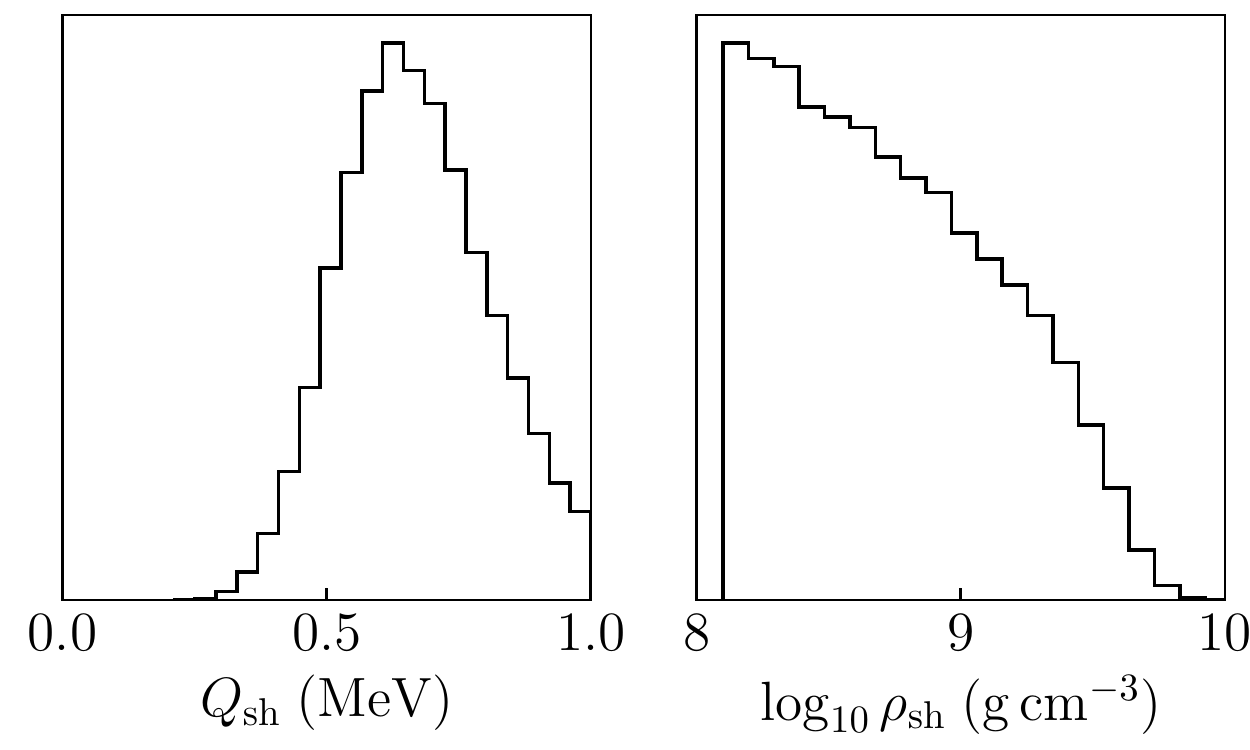}
	\end{center}
	\caption{Histograms of the distribution of the shallow heating strength, $Q_\mathrm{sh}$, and lower density, $\rho_\mathrm{sh}$
	in scenario C.
	}
	\label{fig:ScenarioC_Shallow}
\end{figure}

We also show in \Fig{fig:ScenarioC_Shallow} the posterior distribution of the strength, $Q_\mathrm{sh}$, and lower density, $\rho_\mathrm{sh}$, 
of the shallow heating in our scenario ``C''.
The $Q_\mathrm{sh}$ distribution peaks at 0.6 MeV, which is slightly higher that in scenario ``B'' where the peak was at 0.4 MeV.
As in scenario ``B'' the distribution of $\rho_\mathrm{sh}$ is leaning on its lower permitted value:
exploring this peak at lower densities would, however, imply
lowering the outer boundary density of our models in \texttt{NSCool}
and require a significant extension of the code's numerics that will be implemented in a future paper\footnote{Such an extension has been performed \citep{Beznogov:2020uo} but the resulting code is too slow to be employed in massive calculations as in an MCMC.}.

\begin{figure*}
	\begin{center}
	\includegraphics[width=1.02\textwidth]{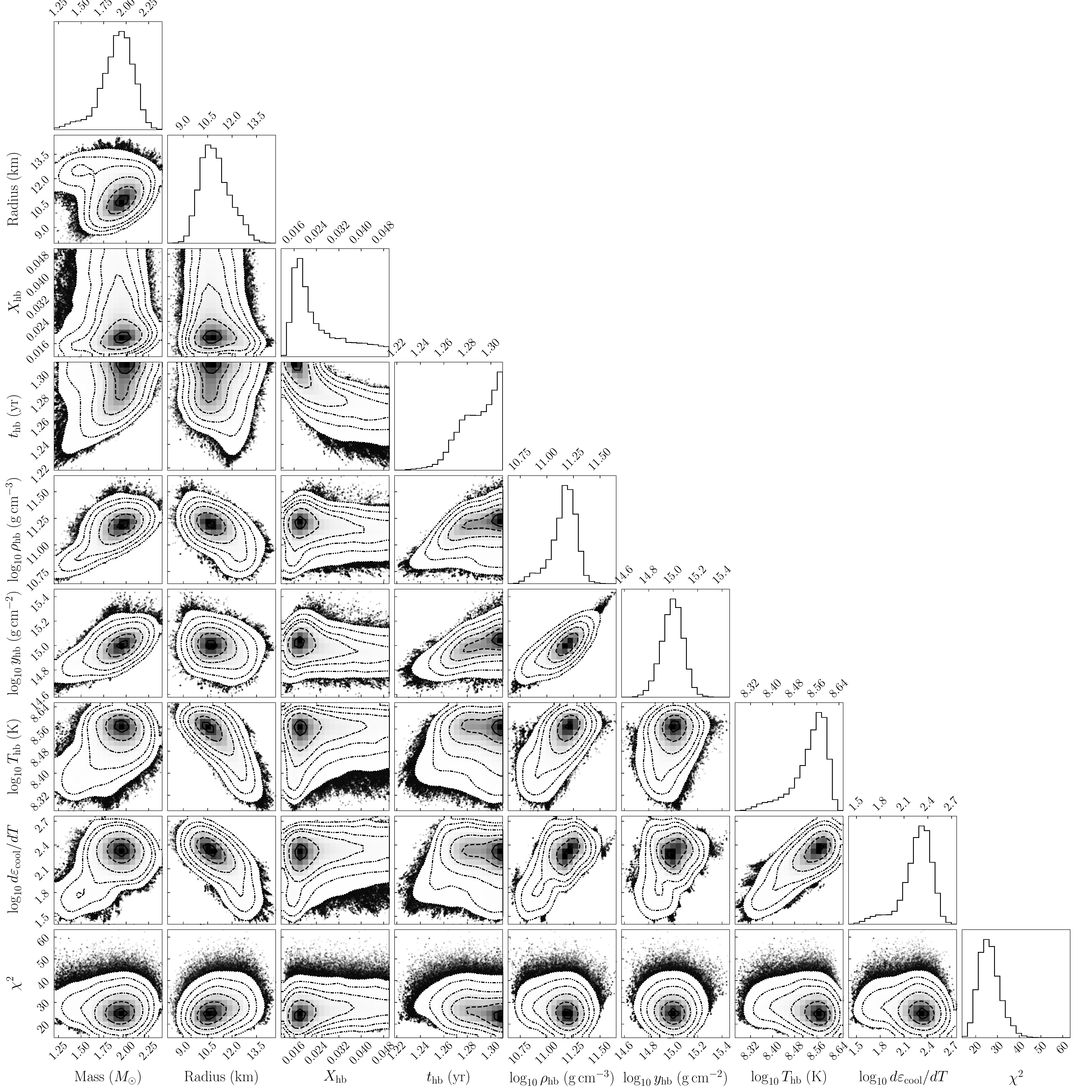}   % this figure has y_{hb}
	\end{center}
	\caption{Corner plot of 1D and 2D histograms of the posterior distributions of five MCMC parameters, mass and radius of the model and the three hyperburst
	control parameters, $X_\mathrm{hb}$, $t_\mathrm{hb}$, and $\rho_\mathrm{hb}$, 
	as well as its equivalent column density $y_\mathrm{hb}$,
	and two deduced quantities, the critical temperature, $T_\mathrm{hb}$,
	and the cooling sensitivity, $\mathrm{log}_{10} \, d\varepsilon_\mathrm{cool}/dT$ 
	(see Eqs. \ref{Eq:Criterion}  and \ref{Eq:eps_cool} with $\varepsilon_\mathrm{cool}$ measured in erg g$^{-1}$ s$^{-1}$), 
	calculated at the critical density at time $t_\mathrm{hb}$,
	as well as the $\chi^2$ of the model fit to the 19 data points.
	}
	\label{fig:CornerC}
\end{figure*}

We emphasize that the MCMC process was driven to fit the data using as parameters for the hyperburst only its occurrence time, $t_\mathrm{hb}$,
depth, $\rho_\mathrm{hb}$, and total energy through $X_\mathrm{hb}$.
We made no further assumption about the nature of the explosion, its fuel and the temperature at which it was triggered.
To characterize the physical conditions of the hyperburst trigger, we display in \Fig{fig:CornerC} a summary of the most important results of this scenario ``C''.
Beside the posterior distribution of 5 of our most relevant MCMC parameters we also report the posterior distribution of two quantities that are output of our calculation:
1) the temperature of the critical layer just before ignition, $T_\mathrm{hb}$, and 2) the cooling sensitivity (Eqs. \ref{Eq:Criterion} and \ref{Eq:eps_cool})
that will be of interest in the next section.
The first noticeable result is a slightly bimodal distribution in terms of mass and radius: the dominant peak is located at $\sim (2 M_{\odot}, 10.5\, \mathrm{km})$, 
and the second smaller one around $\sim (1.5 M_\odot, 12.5\, \mathrm{km})$. 
These peaks also exhibit clear differences in critical density and temperature:
for the first quantity, there is a threshold at $10^{11}$ g cm$^{-3}$ above (below) which the dominant (lower) peak can be found, while for the latter quantity 
their respective temperatures are $\sim 10^{8.6} \approx 4\times 10^8$ K for the dominant and $\sim 10^{8.4} \approx 2.5 \times 10^8$ K for the lower one.
Of relevance also is that the dominant peak favors a trigger time $t_\mathrm{hb}$ very close to the end of the accretion outburst while the second peak
favors $t_\mathrm{hb} \simeq $ 1.25 year, i.e., about 25 days before the end of the outburst.
Notice, finally, that the column density $y_\mathrm{hb}$ corresponding to $\rho_\mathrm{hb}$  has a narrow, symmetric
distribution centered at $10^{15}$ g cm$^{-2}$.

%------------------------------------------------------------------------------------------------
%------------------------------------------------------------------------------------------------

%------------------------------------------------------------------------------------------------
%------------------------------------------------------------------------------------------------

%%%%%%%%%%%%%%%%%%%%%%%%%%%%%%%%%%%%%%%%%%%%%%%%%%%%%%%%%%%%%%%%%%%%%%%%%%%%%%%%
\section{The Fuel of the Hyperburst}
\label{Sec:fuel}
%%%%%%%%%%%%%%%%%%%%%%%%%%%%%%%%%%%%%%%%%%%%%%%%%%%%%%%%%%%%%%%%%%%%%%%%%%%%%%%%

%------------------------------------------------------------------------------------------------
\begin{figure}[t]
	\begin{center}
	\includegraphics[width=0.99\columnwidth]{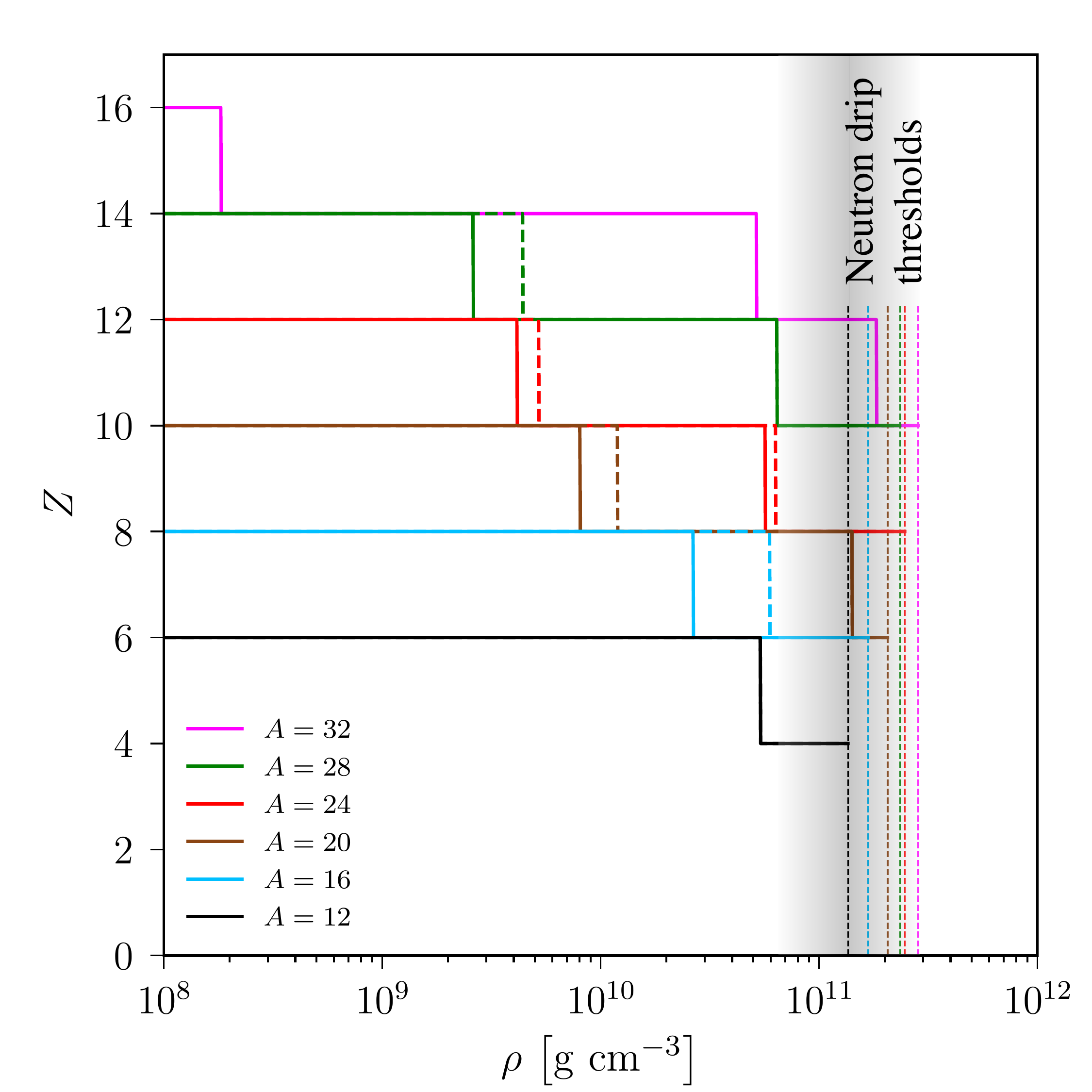}
	\end{center}
	\caption{Evolution of the charge number $Z$, under the effect of double electron captures, of the six lightest 
	$\alpha$-nuclei, $^{12}$C, $^{16}$O, $^{20}$Ne, $^{24}$Mg, $^{28}$Si, and $^{32}$S,
	from their production site, at densities below $10^9$ g cm$^{-3}$, when pushed to increasing densities.
	Continuous lines allow forbidden electron captures while dashed ones only consider allowed transitions.
	Vertical dotted lines show the neutron drip density thresholds for each isobaric sequence. 
	The grey shaded region corresponds to the predicted location of the origin of the hyperburst
	and in this density range excluding or including forbidden transitions makes no difference.
    Details of the calculation are provided in \App{Sec:ecaptures}.}
	\label{fig:ecaptures}
\end{figure}
%------------------------------------------------------------------------------------------------

In the classical short H/He X-ray bursts, as well as the long C superbursts, matter in the stellar envelope is pushed by accretion toward higher densities
and temperatures till it reaches a critical point where the nuclear burning becomes unstable and a thermonuclear explosion ensues.
The time scale for this compression is hours to days in the case of the short bursts and months to years for the superbursts. 
In our scenario, an explosion is triggered at densities $\sim 10^{11}$ g cm$^{-3}$, corresponding to column densities of the order of $10^{15}$ g cm$^{-2}$:
under an Eddington accretion rate ($\dot{m}_\mathrm{Edd} \approx 10^5$ g cm$^{-2}$ s$^{-1}$) at such densities
matter is barely progressing by one millimeter per day and the time scale to reach this point is centuries to millennia.
Moreover, matter is being pushed toward lower temperatures because of the inverted temperature gradient 
(right lower panel of \Fig{fig:Bigbadaboom}).
However, we are not in a steady state situation but rather in a year and a half long strong outburst during which the critical layer is still warming up
(left lower panel of \Fig{fig:Bigbadaboom} and left panel of \Fig{fig:Bigbadaboom2}):
the physical origin of the explosion appears thus to be the result of the temperature at the critical layer to be rising with time rather than this matter
being pushed toward higher densities.

A standard criterion in X-ray burst modeling to identify the location of the ignition layer \citep{fujimoto81,Bildsten:1998wm}
is to compare the temperature dependence of the nuclear burning rate,
$\varepsilon_\mathrm{nucl}$, with the cooling rate
\be
\varepsilon_\mathrm{cool} = \frac{\partial F}{\partial y} = \frac{1}{\rho} \frac{\partial F}{\partial r}
\label{Eq:eps_cool}
\ee
where $F$ is the heat flux.
As long as the inequality
\be
\frac{d \varepsilon_\mathrm{cool}}{dT} > \frac{d \varepsilon_\mathrm{nucl}}{dT}
\label{Eq:Criterion}
\ee
is satisfied the burning is stable.
Histograms of the values of $\mathrm{log}_{10} \, d\varepsilon_\mathrm{cool}/dT$, calculated with the temperature profile of each model
just before the time $t_\mathrm{hb}$ at $\rho_\mathrm{hb}$, were presented in \Fig{fig:CornerC}:
we found that triggering of the hyperburst occurred at values $\mathrm{log}_{10} \, d\varepsilon_\mathrm{cool}/dT \sim 2$
(with $\varepsilon_\mathrm{cool}$ in erg g$^{-1}$ s$^{-1}$).

The candidate nucleus to trigger the hyperburst must have been produced by nuclear burning, either stable or explosive, 
at lower densities and then pushed to high densities by accretion.
On its journey this nucleus will undergo electron captures as described, e.g., by \cite{Haensel:1990kx}.
The most abundant light nuclei produced by the nuclear burning in the envelope are $\alpha$-nuclei and in \Fig{fig:ecaptures} we show how the
six lightest ones (excluding $^{8}$Be which is unstable) evolve under double electron captures when pushed to increasing densities:
these are our pre-candidates for the triggering of the hyperburst.

%------------------------------------------------------------------------------------------------
\begin{figure*}[t]
	\begin{center}
	\includegraphics[width=0.99\columnwidth]{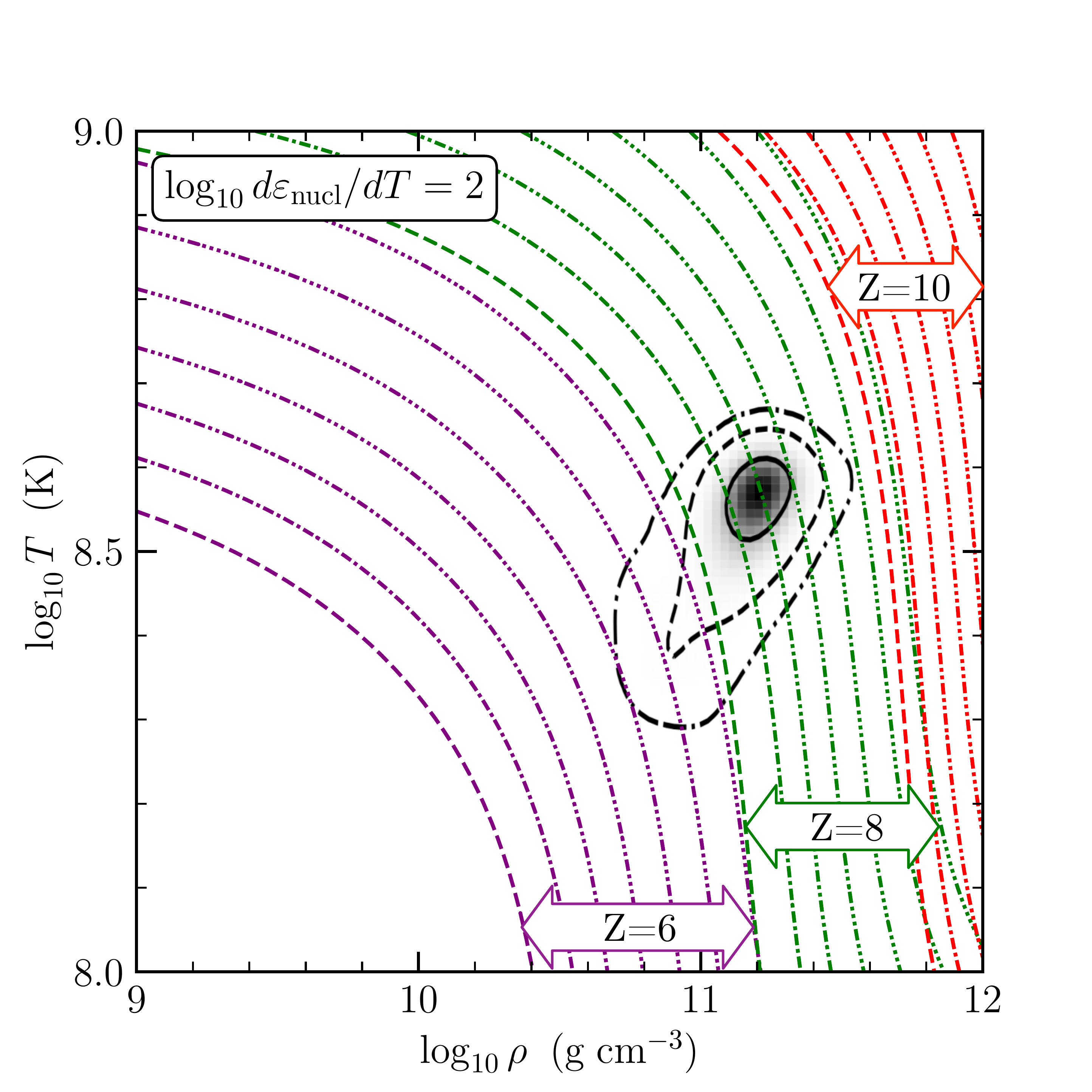}
	\includegraphics[width=0.99\columnwidth]{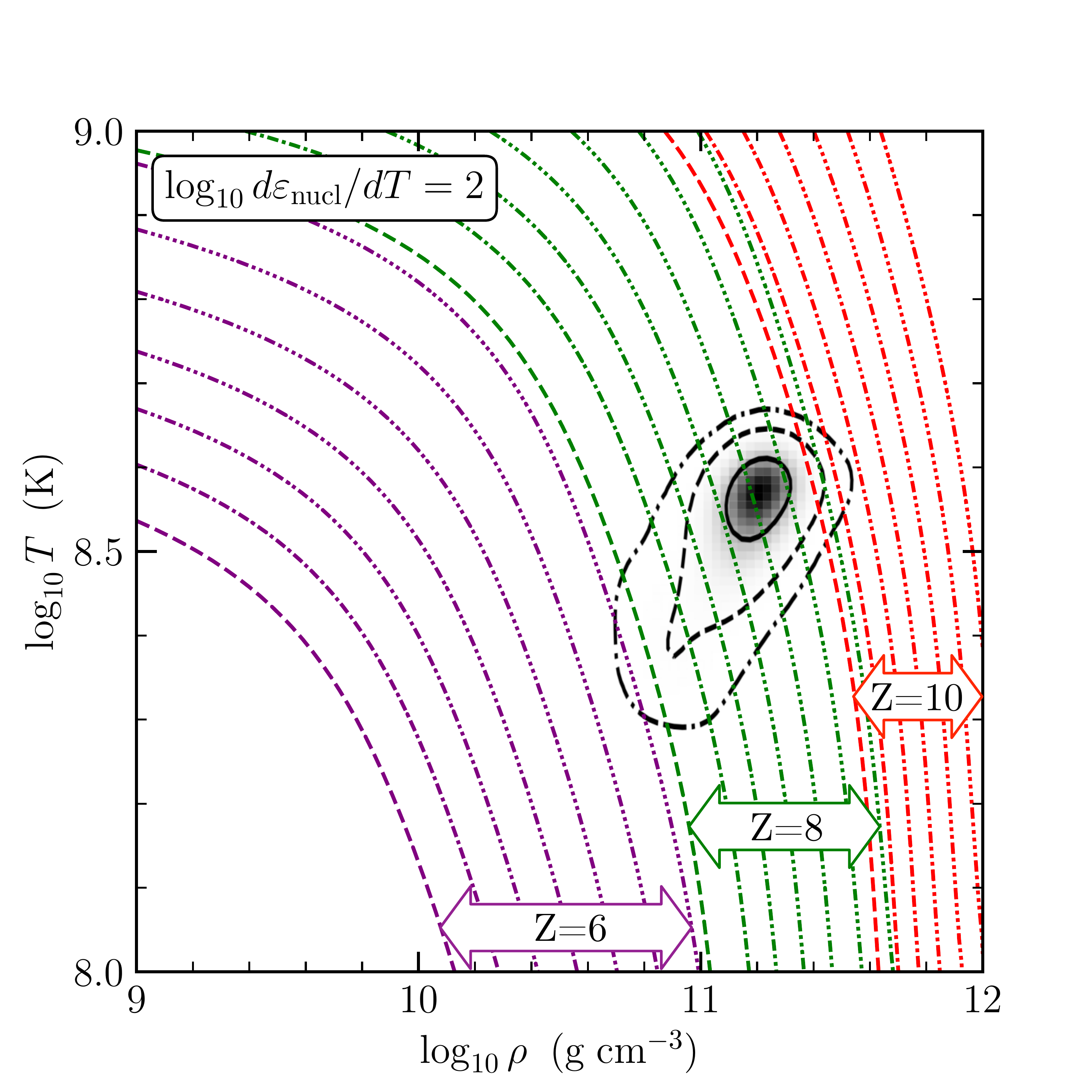}

	\includegraphics[width=0.99\columnwidth]{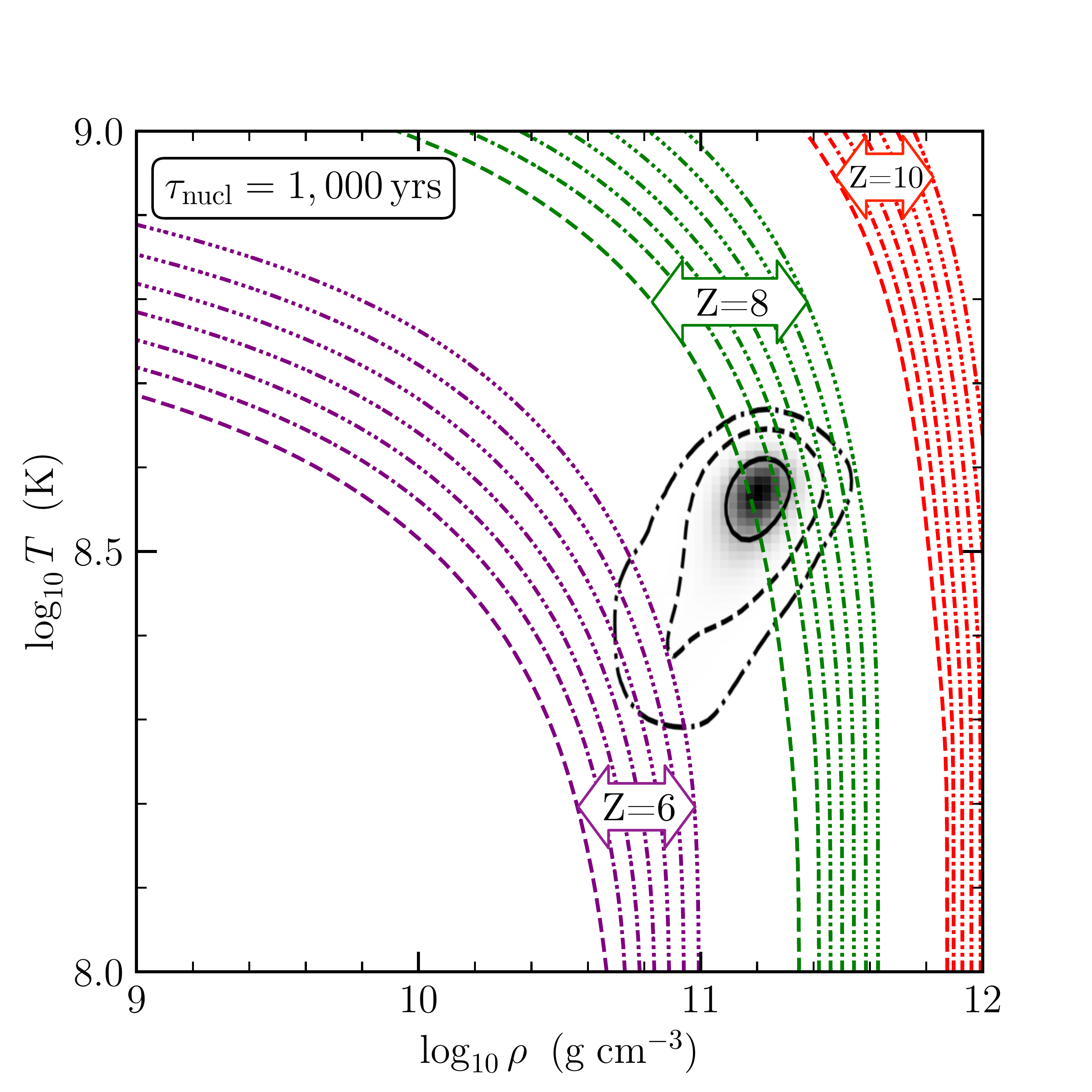}
	\includegraphics[width=0.99\columnwidth]{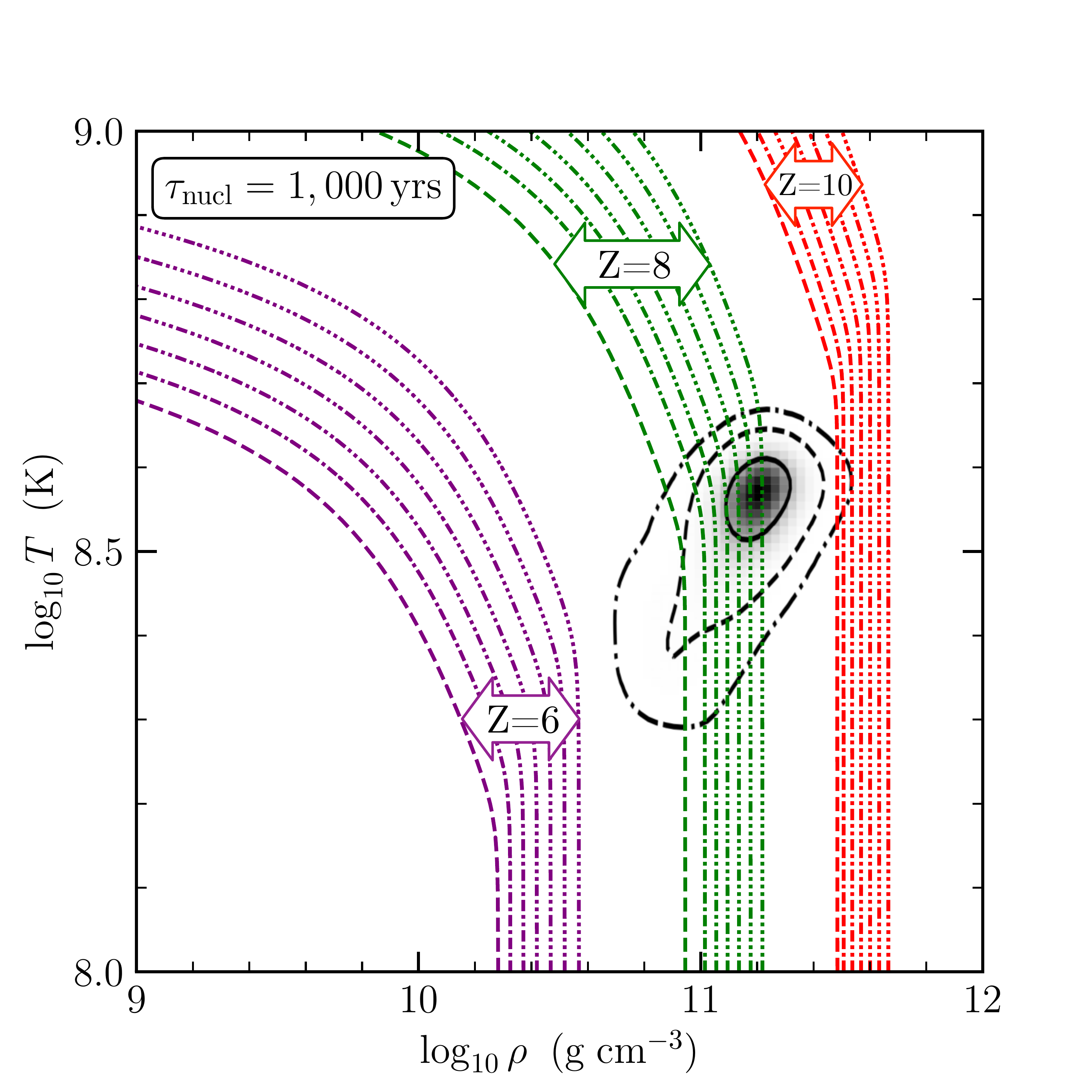}

	\end{center}
	\caption{
	Upper panels: contour lines where $\mathrm{log}_{10} d\varepsilon_\mathrm{nuc}/dT = 2$ ($\varepsilon_\mathrm{nucl}$ being in erg g$^{-1}$ s$^{-1}$ and $T$ in K)
	for the fusion of $^{16}$C+$^{16}$C (labelled ``$Z=6$''), $^{20}$O+$^{20}$O (labelled ``$Z=8$''), and $^{28}$Ne+$^{28}$Ne (labelled ``$Z=10$''),
	and for each case with seven mass fractions $X=10^{-n}$  with $n=0, 1, 2, \dots ,6$, as dash-$n$-dot lines.
	Lower panels: contour lines of life-time $\tau=10^3$ yrs of the $^{16}$C, $^{20}$O, and $^{28}$Ne nuclei 
	under the same fusion reactions as in the upper panels.
	For the treatment of screening, these nuclei are assumed immersed in a background of $^{56}$Ca,
	with mass fraction $1-X$, which is the
	nuclear species predicted to be present at $\rho \sim 10^{11}$ g cm$^{-3}$ in the model of \citet{Haensel:2008aa}.
        In the left panels we apply the minimum reaction rates while in the right panel the maximum ones are used.
	Details of the calculations are presented in \App{Sec:Nucl}.
	In all panels the background contours reproduce the $1\sigma$, $2\sigma$, and $3\sigma$, confidence ranges of the 2D distribution
	of $\mathrm{log}_{10} \, \rho_\mathrm{hb} - \mathrm{log}_{10} \, T_\mathrm{hb}$ already displayed in \Fig{fig:CornerC}.
	}
	\label{fig:fusion}
\end{figure*}
%------------------------------------------------------------------------------------------------

One interesting feature of the physical conditions where the hyperburst is triggered, which may affect the conclusions about the triggering species, is that the crust main matter, in the temperature and density ranges where the explosion starts, is in a solid (likely crystalized) state: the lower right panel of \Fig{fig:Bigbadaboom} displays the melting temperature curve of the main matter which shows that on this temperature profile solidification occurs close to $10^{10}$ g cm$^{-3}$. 
However, it has been predicted \citep{Horowitz:2007tu} that when the high $Z$ nuclei solidify there is a phase separation with the low $Z$ ones that remain in the liquid phase: the emerging state could thus be one of bubbles of liquid O and/or Ne embedded in a solid medium. We also display in the lower right panel of \Fig{fig:Bigbadaboom} the melting temperature for $^{28}$Ne bubbles in these conditions which shows that they remain in a liquid state, given this temperature profile, up to densities well above $10^{11}$ g cm$^{-3}$. Oxygen bubbles have even lower melting temperatures than neon ones because of their lower $Z$.

In \Fig{fig:fusion} we present the critical conditions for the burning of $\alpha$-nuclei with $Z=6, 8$, and 10 (details are presented in \App{Sec:Nucl}) in the $\rho - T$ plane where
the $1\sigma$, $2\sigma$, and $3\sigma$, critical confidence regions for the values of $(\rho_\mathrm{hb}, T_\mathrm{hb})$ from \Fig{fig:CornerC} are shown in the background.
In the upper panels we plot contour lines of $\mathrm{log}_{10} \, d\varepsilon_\mathrm{nucl}/dT = 2$, since $\mathrm{log}_{10} \, d\varepsilon_\mathrm{cool}/dT = 2$ is a typical value found, see \Fig{fig:CornerC}, in our scenario ``C'' MCMC.
In the lower panels we plot contour lines of the lifetimes $\tau = 1,000$ yrs of the same nuclei, since 1,000 yrs is a typical time for matter to be pushed to these densities by accretion.
Even if the medium average mass fraction $\langle X \rangle$ of the triggering nucleus is small, its mass fraction in the bubbles where it is concentrated during medium crystalization, if concentration did occur, could be close to unity.
To encompass this range of possibilities 
all critical contour lines in \Fig{fig:fusion} consider a range of $X$ from $10^{-6}$, no bubble formation, up to unity, perfect bubble formation.
Depending on the actual mass fraction of the possible exploding nuclei we have a variation of a factor of a few in the predicted density where the explosion started.
At these densities and relatively low temperatures, uncertainties on the burning rates are largely dominated by uncertainties in the screening and we apply either the estimated minimum rates, in the left panels, or the maximum ones, in the right panels.
Notice that the burning rates depend strongly on the charge number, $Z$, of the nuclei and have much less sensitivity to their mass number $A$ so that considering different isotopes has little impact on this part of our inquiry.
From this \Fig{fig:fusion} we see that, clearly, carbon is not the trigger nucleus of our hyperburst (it is most likely the trigger nucleus of superbursts acting at lower densities $\sim 10^{9}$ g cm$^{-3}$, \citealt{Cumming:2001vy} and \citealt{Cumming:2006uv}): it is only at very low concentrations, $X \ll 10^{-4}$, that carbon burning can be unstable within the critical region of the $\rho - T$ plane but, unless something beyond our simple calculations happened, it would have been consumed by stable burning before reaching such high densities as seen from the lower panels of \Fig{fig:fusion}.
Oxygen appears to be the best candidate but neon cannot be excluded particularly if actual reaction rates are close to the maximum value we apply and the hyperburst critical density was on the high values found by the MCMC. Oxygen could be $^{20}$O coming from $^{20}$Ne after a pair of electron captures or $^{24}$O coming from $^{24}$Mg after two pairs of electron captures, as pictured in \Fig{fig:ecaptures}. Similarly, neon could be $^{28}$Ne coming from $^{28}$Si or $^{32}$Ne coming from $^{32}$S.
The lower panels of \Fig{fig:fusion} show that both oxygen and neon nuclei need to be pushed to densities higher than their explosion densities before they are exhausted and, thus, exhaustion by stable burning is not an issue.

%%%%%%%%%%%%%%%%%%%%%%%%%%%%%%%%%%%%%%%%%%%%%%%%%%%%%%%%%%%%%%%%%%%%%%%%%%%%%%%%
\section{Discussion}
\label{Sec:discussion}
%%%%%%%%%%%%%%%%%%%%%%%%%%%%%%%%%%%%%%%%%%%%%%%%%%%%%%%%%%%%%%%%%%%%%%%%%%%%%%%%

Having now data for a fourth outburst in \MAXI, our first finding is that shallow heating during outbursts 2, 3, and 4 can be described by identical parameters,
$Q_\mathrm{sh}$ and $\rho_\mathrm{sh}$ (see Eq. \ref{Eq:Heating} and the description following it).
The thermal evolution of this neutron star when exiting its first outburst implies, however, a very different physical condition,
as is easily intuited by a look at \Fig{fig:Cool_All}.
Such condition had previously been described as an extremely strong continuous shallow heating, with $Q_\mathrm{sh} \sim 10$--$20$ MeV compared to 
$\sim 1$ MeV in outbursts 2 and 3  \citep{Deibel:2015aa,Parikh:2017uj}.

Our second finding is that the high temperature of the \MAXI\ neutron star when exiting outburst 1, as well as its subsequent years long thermal evolution, can be
very well modeled assuming a very strong and sudden energy release, a ``hyperburst'', instead of a continuous energy release as in the shallow heating scenario.
From our extensive MCMC simulation, scenario ``C'', we find that this event would have occurred during the last 3 weeks of the accretion outburst and released an energy 
of the order of $10^{44}$ ergs.
In this scenario the energy was deposited in a layer extending from the lowest density up to $\sim 10^{11}$g cm$^{-3}$:
global energetics as well as other characteristics are presented in Figs. \ref{fig:ScenarioC_Energy} and \ref{fig:CornerC}.
Shallow heating was also assumed to occur during the first outburst, and described by the same parameters as in the next three outbursts
(see \Fig{fig:ScenarioC_Shallow} for the parameter posterior distributions).

To identify candidates for the triggering of the hyperburst we simply applied the standard one-zone model criterion of \Eq{Eq:Criterion}.
Values of $d \varepsilon_\mathrm{cool}/dT$ were obtained directly from perturbation of the $T$-profile in each one the models of our MCMC and we found that
$\mathrm{log}_{10 }d \varepsilon_\mathrm{cool}/dT \sim 2$, see \Fig{fig:CornerC}, with $\varepsilon_\mathrm{cool}$ measured in erg g$^{-1}$ s$^{-1}$.
Values of $\mathrm{log}_{10 }d \varepsilon_\mathrm{nucl}/dT = 2$ were calculated and displayed in the two upper panels of \Fig{fig:fusion}:
considering the enormous uncertainties on the fusion rate due to screening in, and near to, the pycnonuclear regime, the left upper panel employed minimum
rates and the right upper panel maximum ones.
Lower panels of the same figure show  ``exhaustion lines'', i.e., contour lines where the burning time scale is 1,000 yrs, implying that the described nucleus
would be depleted by stable burning when reaching the corresponding density since 1,000 yrs is roughly the time it takes, under {\MAXI}'s average mass accretion rate,
for matter to reach $10^{11}$ g cm$^{-3}$.
The conclusion of this is that C could be the triggering nucleus, but it is unlikely to be able to reach the required density before being depleted,
leaving O and Ne as the best candidates.
Isotopes of O and Ne have slightly different burning rates, but differences are small enough that the results of \Fig{fig:fusion} are practically only dependent on
the charge $Z$ of the nucleus and not on their mass number $A$.
Once produced at low densities, nuclei are compressed by accretion and undergo double electron capture reactions that gradually reduce their charge $Z$,  see \Fig{fig:ecaptures},
but without changing the mass $A$:
the triggering oxygen is likely $^{20}$O or even $^{24}$O, descendants of $^{20}$Ne and $^{24}$Mg,
while the triggering neon would be $^{28}$Ne or $^{32}$Ne, descendants of $^{28}$Si and $^{32}$S.

Fusion of low $Z$ nuclei liberates about 1 MeV per nucleon (see \Tab{tb:Q}), i.e. about $10^{18}$ erg g$^{-1}$,
and we parametrized the energy injected during the explosion as $X_\mathrm{hb} \times 10^{18}$ erg g$^{-1}$ 
so that $X_\mathrm{hb}$ approximately reflects the average mass fraction, $\langle X \rangle$, of the exploding nuclei.
Our MCMC found a peak at $X_\mathrm{hb} \sim 0.02$, exhibited in \Fig{fig:CornerC}, while the minimum value found was 0.011.
Thus, a small mass fraction of a few percents of low $Z$ nuclei is needed to provide the required energy.
While accretion is ongoing, in the upper layers of the accreted material H burns into He through the hot CNO cycle and He into C
through the triple-$\alpha$ reaction.
At  high accretion rates, as is the case in \MAXI, breakout reactions from the CNO cycle add the rp-process  \citep{Wallace:1981ud} to H burning 
and lead to the production of high $Z$ nuclei, possibly up to the SnSbTe cycle at $Z=50$--$52$ and $A=103$--$107$ \citep{Schatz:2001vm}.
In the absence of X-ray bursts, the models of \citet{Schatz:1999us} with continuous burning in the ocean at accretion rates $\sim \dot{M}_\mathrm{Edd}$  produced 
$A=12$ nuclei at a mass fraction $\sim 4$\% and $A=24$ at $\sim 2-3$\% with other low $A$ nuclei in much smaller abundances and heavy nuclei up to $A \sim 80$
through the rp-process.
This is likely representative of the burning in most of the observed accretion phases of \MAXI\ that were close to $\dot{M}_\mathrm{Edd}$ most of the time
(see, e.g., upper panel of \Fig{fig:Bigbadaboom}) as explosive burning is quenched at high mass accretion rates \citep[e.g.][]{Bildsten:1998wm,Galloway:2021vr}.
However, during low mass accretion moments, as, e.g., toward the end of an accretion outburst, bursting behavior may appear as was observed in the case
of the end of the outburst of XTE J1701-462 when $\dot{M}$ was down to about 10\% $\dot{M}_\mathrm{Edd}$ \citep{Lin:2009vd}.
We notice that \MAXI's first outburst, in 2011-2012, was almost a carbon copy of XTE J1701-462's 2006-2007 outburst but, nevertheless, no X-ray bursts were detected
while we may have found a small X-ray burst toward the end of the 2020 outburst as presented in \S~\ref{sec:nicer}.
Within this context of bursting behaviour, we can consider the ashes of the three cases used by \citet{Lau:2018tr}:
in the extreme rp-process X-ray burst case (based on \citealt{Schatz:2001vm}) 
one finds negligible amounts of C and O left but about 1\% of $A=20$ nuclei, 
while in their \texttt{KEPLER} X-ray burst case (based on \citealt{cyburt2016}) one has to go up to $A=28$ and $32$ to find nuclei with a significant mass fraction, about 5\% and 10\%, respectively:
in both cases we have enough seed nuclei which, after multiple double electron captures when pushed to high densities,
could be the trigger nuclei for a hyperburst.
The third case we contemplate is the occurrence of a superburst (made possible, e.g., by C ashes from long term continuous burning as seen above):
the resulting ashes (based on  \citealt{keekheger2011}) contain negligible amount of C, O, and Ne, but almost 1\% of $A=28$ nuclei which could lead to a seed nucleus for
hyperburst triggering after electron captures.
Since the hyperburst is triggered at densities $\sim 10^{11}$ g cm$^{-3}$, a region whose chemical composition is the result of many centuries of accretion,
of which we only have a decade long snapshot, all of the above burning possibilities may be partial realities and show that
having $\sim 1-3$\% of low $Z$ nuclei present at such densities is a realistic possibility.

We see from \Fig{fig:CornerC} that values of the column density at the explosion point, $y_\mathrm{hb}$, range from $5\times 10^{14}$ up to $2\times 10^{15}$ g cm$^{-2}$.
Under an Eddington rate it takes then approximately from 150 to 600 years, respectively, 
of continuous accretion for matter to be pushed to the explosion point.
From \MAXI's inferred bolometric flux and \Eq{Eq:Mdot} we deduce a total accreted mass, in the 
four observed outbursts, of almost $10^{26}$ g resulting in a 10 years average
$\langle \dot M \rangle \simeq 2.8 \times 10^{17}$ g s$^{-1} \simeq 4.4\times 10^{-9} M_\odot$ yr$^{-1}$, i.e., about 30\% of the Eddington rate.
Since \MAXI\ was never seen previously to its 2011-2012 outburst the long term actual fraction of time it spends in quiescence is possibly much larger than indicated by the last decade of observations
and its long term $\langle \dot M \rangle$ smaller that estimated above.
Hyperbursts in \MAXI\ are thus likely a once in a few millennia events and the probability to have witnessed one in 10 years since its discovery, or in 25 years of having continuous all sky monitoring in X-rays, is of the order of 1\%: a small but not too small value.
The seven known persistent Z sources \citep{hasinger1989,fridriksson2015}, also estimated to be accreting close to, or above, the Eddington limit most of the time, are likely to experience hyperbursts more frequently, once every few centuries, but unfortunately these are not detectable under continuous accretion due to the small increase in surface luminosity during the explosion, see panel (f) in \Fig{fig:ScenarioC_Energy}.
From what we found in the present work, a hyperburst can only be recognized if accretion stops
shortly afterwards and we detect an anomalously hot neutron star.
It could be considered too much of a coincidence that the hyperburst need to have occurred only a few days/weeks before the end of the outburst.
Notice, however, that even if it had happened several months earlier, the star would still have exited the outburst with a very high effective temperature and allowed the identification of the
occurrence of the hyperburst: \Fig{fig:Cool_All} shows that it took more than 100 days after the end of the outburst, i.e., after the occurrence of the hyperburst, for the crust to cool down and $T_\mathrm{eff}^\infty$ to drop below 250 eVs.
Had the hyperburst occurred 3-4  months before the end of the outburst, the star would have entered quiescence with a temperature in excess of 250 eVs which would already have singled out \MAXI\ as an anomalously hot star.
At least in the case of \MAXI\, Nature appears to have been cooperating!

We now have a handful of binary systems under transient accretion in which post outburst cooling data are available. 
Most of them have much lower mass accretion rates than \MAXI\, which implies that their crust 
experience lower temperatures and that their matter would need to be compressed to still higher densities for a hyperburst to be triggered.
Hyperbursts in these system are, thus, even rarer events than in high accretors and
occurrence, and identification, of one in any of these systems appears to be, unfortunately, highly unlikely, unless many more such systems are discovered.
XTE J1701-462 is the exception within this sample as it exhibited an outburst very similar to \MAXI\ first observed outburst, while its neutron star exited it with a much lower surface temperature of the order of 150 eV: it may need many more such outbursts till a hyperburst is triggered.

%%%%%%%%%%%%%%%%%%%%%%%%%%%%%%%%%%%%%%%%%%%%%%%%%%%%%%%%%%%%%%%%%%%%%%%%%%%%%%%%
\section{Conclusions}
\label{Sec:conclusions}
%%%%%%%%%%%%%%%%%%%%%%%%%%%%%%%%%%%%%%%%%%%%%%%%%%%%%%%%%%%%%%%%%%%%%%%%%%%%%%%%

We have shown that the highly anomalous thermal evolution of the \MAXI\ neutron star when exiting its 2011-2012 accretion outburst and during the following years
can be very well modeled as due to the occurrence of a hyperburst, i.e., explosive nuclear burning of some neutron rich isotope of O or Ne in a region of density
$\sim 10^{11}$  g cm$^{-3}$. 
Such a deep explosion, depositing $\sim 10^{44}$ ergs, results in a hot outer crust whose cooling takes several years and provides an excellent fit to the data.
Published models of nuclear, either stable or explosive, burning during accretion show that the subsistence of a small amount, a few percents in mass fraction,
of the needed low $Z$ nuclei to these densities is likely.
However, it takes many centuries of accretion to accumulate this material in transient systems and hyperbursts as postulated here are, thus, very rare events 
and we are unlikely, unfortunately, to witness another one. 

Our complementary results that modeling of the cooling of the neutron star after the subsequent three accretion outbursts, in 2012, 2015-2016, and 2020,
can be realized employing a single parametrization of the still enigmatic shallow heating, lead to a prediction about the future cooling of the 
neutron star which will be monitored in the coming years.
The example presented in \Fig{fig:Bigbadaboom} shows a future smooth evolution, see the "Post-outburst 4'' panel, for at least 1,000 days, and the
$3\sigma$ range deduced from our scenario ``C'' MCMC run (containing more than 4 millions cooling curves) shows that this simple behavior is a strong prediction 
of the scenario.

In the global modeling of 12 years of evolution of \MAXI\ in our scenario ``C'' with the hyperburst occurring at the end of the first outburst, parameters describing the
shallow heating during this outburst were also assumed to have identical values as in the next three outbursts.
We can, hence, describe the whole evolution of the \MAXI\ neutron star during a period of 12 years through 4 very different accretion outbursts with a single consistent
parametrization of the shallow heating.
This result is in agreement with the study of the MXB 1659-29 neutron star that could describe its evolution through two accretion outbursts, spanning a period of 
almost two decades (1999 till 2018), also with a single shallow heating parametrization \citep{Parikh:2019aa}.
These results are, however, in contrast with the modeling of the evolution of the Aquila X-1 neutron star, which exhibits frequent but short accretion outbursts,
that required a variation of the shallow heating parameters between different outbursts \citep{Degenaar:2019aa}.

%%%%%%%%%%%%%%%%%%%%%%%%%%%%%%%%%%%%%%%%%%%%%%%%%%%%%%%%%%%%%%%%%%%%%%%%%%%%%
\begin{acknowledgments}
We would like to thank the {\it NICER} and {\it Swift} teams for approving and executing the observations made during and after the 2020 outburst.
DP and MB acknowledge financial support by the Mexican Consejo Nacional de Ciencia y Tecnolog{\'\i}a with a CB-2014-1 grant $\#$240512 and the Universidad Nacional Aut\'onoma de M\'exico through an UNAM-PAPIIT grant \#109520.
MB also acknowledges support from a postdoctoral fellowship from UNAM-DGAPA and support from a grant of the Ministry of Research, Innovation and Digitization, CNCS/CCCDI -- UEFISCDI, Project No. PN-III-P4-ID-PCE-2020-0293, within PNCDI III. This research has made use of data and/or software provided by the High Energy Astrophysics Science Archive Research Center (HEASARC), which is a service of the Astrophysics Science Division at NASA/GSFC. This work made use of data supplied by the UK Swift Science Data Centre at the University of Leicester.
\end{acknowledgments}

\facilities{CXO, NICER, Swift, XMM}

\software{NSCool \citep{Page:2016aa}, Numpy \citep{Numpy}, Matplotlib \citep{Matplotlib}, HEASOFT \citep{ftools}} 

%%%%%%%%%%%%%%%%%%%%%%%%%%%%%%%%%%%%%%%%%%%%%%%%%%%%%%%%%%%%%%%%%%%%%%%%%%%%%

%%%%%%%%%%%%%%%%%%%%%%%%%%%%%%%%%%%%%%%%%%%%%%%%%%%%%%%%%%%%%%%%%%%%%%%%%%%%%%%%
%%%%%%%%%%%%%%%%%%%%%%%%%%%%%%%%%%%%%%%%%%%%%%%%%%%%%%%%%%%%%%%%%%%%%%%%%%%%%%%%
\appendix
%%%%%%%%%%%%%%%%%%%%%%%%%%%%%%%%%%%%%%%%%%%%%%%%%%%%%%%%%%%%%%%%%%%%%%%%%%%%%%%%
%%%%%%%%%%%%%%%%%%%%%%%%%%%%%%%%%%%%%%%%%%%%%%%%%%%%%%%%%%%%%%%%%%%%%%%%%%%%%%%%

%%%%%%%%%%%%%%%%%%%%%%%%%%%%%%%%%%%%%%%%%%%%%%%%%%%%%%%%%%%%%%%%%%%%%%%%%%%%%%%%
\section{Some details on the crust physics we employ}
\label{Sec:physics}
%%%%%%%%%%%%%%%%%%%%%%%%%%%%%%%%%%%%%%%%%%%%%%%%%%%%%%%%%%%%%%%%%%%%%%%%%%%%%%%%

We take the equation of state and chemical composition of the crust from \citet{Haensel:2008aa} 
for the model assuming $^{56}$Fe at low densities, and fix the crust-core transition density
at $\rho_\mathrm{cc} \approx 1.5\times 10^{14}$ g cm$^{-3}$. 
Reasonable variation of this transition density is known to have very little effect on the crust relaxation modeling \citep{Lalit:2019aa}.

The specific heat is obtained by adding the contribution of the degenerate gas of electrons,
the degenerate gas of dripped neutrons in the inner crust and the nuclei, including the Coulomb interaction contribution
in the liquid phase from \citet{Slattery:1982aa}  and in the solid phase from \citet{Baiko:2001aa} 
with a liquid-solid phase transition occurring at a Coulomb coupling parameter $\Gamma = 180$.
The only strong interaction modifications to the neutron specific heat we include is from the effect 
of pairing for which we follow \citet{Levenfish:1994aa} while the pairing phase transition 
critical temperature is taken from  \citet{Schwenk:2003aa}.

For the thermal conductivity, dominated by electrons, we follow \citet{Potekhin:1999aa} and
\citet{gnedin:2001aa} for electron-phonon scattering in the solid phase to which we add
electron-impurity scattering following the simple treatment of \citet{Yakovlev:1980aa}.
When ions are in the liquid phase we apply the results of \citet{Yakovlev:1980aa}.
We neglect the very small contribution of electron-electron scattering \citep{shternin:2006aa}.
We also neglect heat transport from neutrons in the inner crust that is a very minor 
contribution \citep{schmitt:2018aa}.

In a very hot crust as in \MAXI, neutrino emission from plasmon decay can become a significant
sink of energy and we employ the result of \citet{itoh:1996aa}.
We also include neutrino emission from pair annihilation from the same authors, a process
that may contribute in  cases of extremely hot outer crusts.
For the small contribution of electron-ion bremsstrahlung we follow \citet{Kaminker:1999aa}
and we also include neutrino emission for the Cooper breaking and formation process in $^1$S$_0$ neutron 
superfluid phase transition in the inner crust following the treatment described in \citet{Page:2009aa}.
We do not include neutrino emission from the Urca cycles \citep{Schatz:2014vo} which may have a significant effect \citep{Deibel:2015aa}
but would force us to introduce more parameters in our MCMC runs to explore the variations in chemical composition that control it.

Our handling of the heating, both from deep crustal heating and shallow heating, and the thermonuclear
explosion was described in \S~\ref{Sec:Modeling}.
We neglect the effect of nuclear burning in the envelope after the accretion phase, in the form of diffusive nuclear burning, as it has only a small effect for a few days \citep{Wijngaarden:2020aa}.

We do not focus on the physics of the stellar core in the present work and we simply follow
the minimal scenario as described in \citet{Page:2004uh} with no pairing taken into account.

%%%%%%%%%%%%%%%%%%%%%%%%%%%%%%%%%%%%%%%%%%%%%%%%%%%%%%%%%%%%%%%%%%%%%%%%%%%%%%%%
\section{Some Details of our Monte Carlo Runs}
\label{Sec:MCMC}
%%%%%%%%%%%%%%%%%%%%%%%%%%%%%%%%%%%%%%%%%%%%%%%%%%%%%%%%%%%%%%%%%%%%%%%%%%%%%%%%

\begin{figure}
	\begin{center}
	\includegraphics[width=0.4\textwidth]{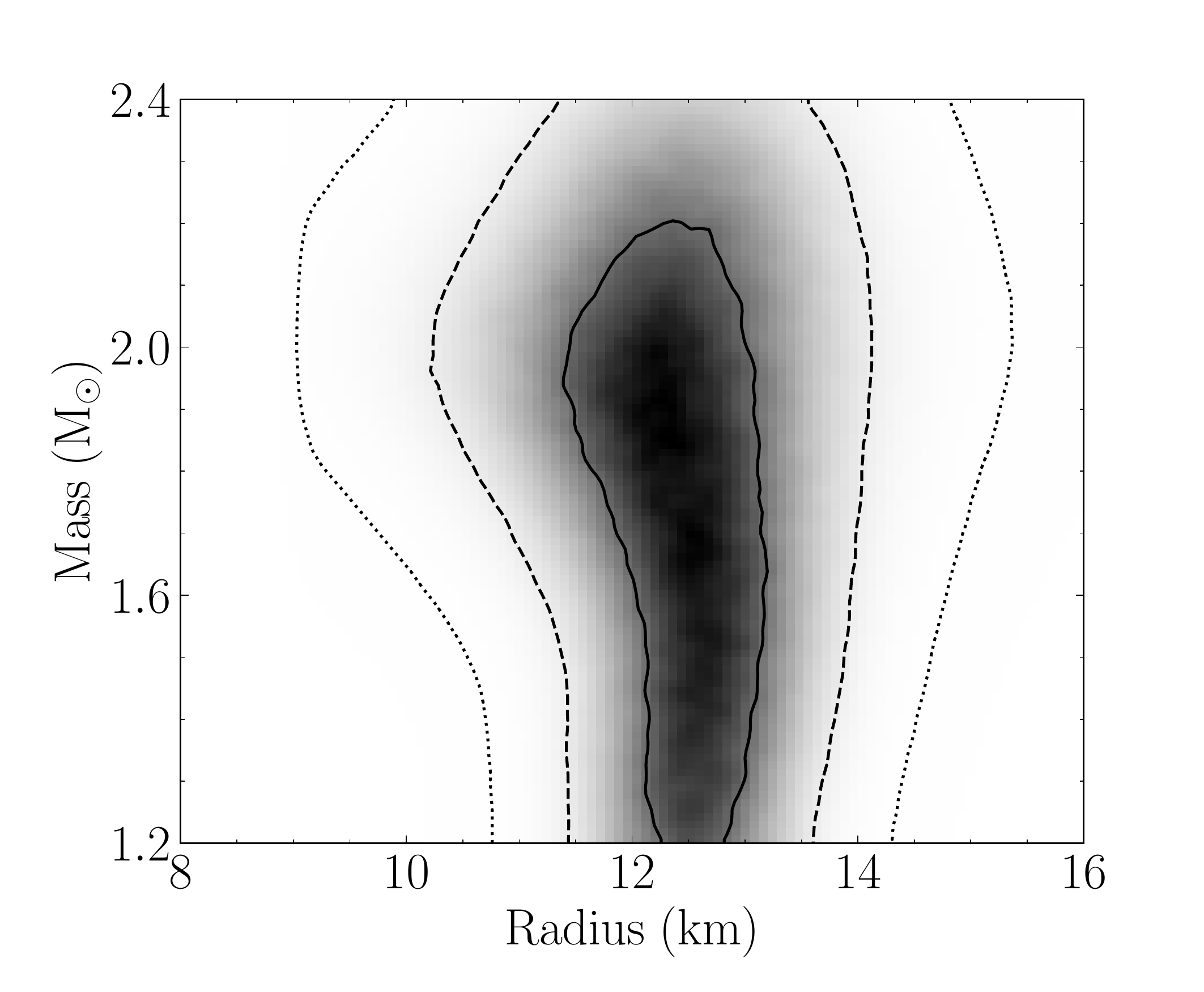}
	\end{center}
	\caption{Our prior for the mass and radius distribution. The three contours, continuous, dashed and dotted,
	show the $1\sigma$, $2\sigma$, and $3\sigma$, ranges, respectively.
	}
	\label{fig:M-R}
\end{figure}

Our priors for the distributions of the MCMC parameters are as follows.
For the mass $M$ and radius $R$ of the neutron stars we limit the range of the former from 
$1.2$ to $2.4$ $M_\odot$ and of the latter from 8 to 16 km with a joint probability distribution 
in these ranges that is displayed in \Fig{fig:M-R}.
This distribution encompasses the deduced posteriors from the two classes of models, ``PP'' and ``CS'', of \citet{Raaijmakers:2021aa}.
The initial, red-shifted, core temperature $\widetilde{T}_0$ is taken as uniformly distributed in the range
$(0.1 - 1.5)\times 10^8$ K.
About the column density of light elements in the envelope, $y_\mathrm{L}$, measured in g cm$^{-2}$, 
we assume a uniform distribution of $\mathrm{log}_{10} \, y_\mathrm{L}$ in the range 6 to 10,
and choose it independently at the beginning of each accretion outburst, thus having four values 
$y_\mathrm{L}(k)$, $k=1, \dots,4$.
The resulting $\widetilde{T}_0 - T_\mathrm{eff}^\infty$ relationship is illustrated in the right panel of
Figure 3 in \citet{Degenaar:2017aa}.
Our next set of parameters are the impurity parameters, $Q_\mathrm{imp}$, 
and we divide the crust in five density ranges, in g cm$^{-3}$:
$[10^8, 10^{11}]$, $[10^{11}, 10^{12}]$, $[10^{12}, 10^{13}]$, $[10^{13}, 10^{14}]$, $[10^{14}, \rho_\mathrm{cc}]$, with a value $Q_\mathrm{imp}^{(i)}$, $i=1, \cdots, 5$, in each range that is uniformly distributed between 0 and 100.
The parametrization of the heating, either shallow or from the hyperburst has been described in the main text
in \Sec{Sec:Modeling}.

When varying $M$ and $R$ we recalculate the structure of the crust by integrating the Tolman-Oppenheimer-Volkoff equation of hydrostatic equilibrium
from the outer boundary at $\rho = \rho_b = 10^8$ g cm$^{-3}$ and radius $r=R$ inward till we reach the crust core density $\rho_\mathrm{cc}$,
thus giving us the core's mass and radius, $m_c$ and $r_c$.
This procedure gives us a self-consistent structure of the crust.
In this work, we are not interested in the response of the core but still need to define its density and chemical composition profile to employ \texttt{NSCool}.
For this purpose we start with a core structure calculated using the APR EOS \citep{Akmal:1998aa} and a stellar mass of $1.4 M_\odot$
(that has a core mass $m_c^\mathrm{(1.4)} = 1.37 M_\odot$ and radius $r_c^{(1.4)} = 10.6$ km)
which gives us the density and mass profiles $\rho = \rho^{(1.4)}(r)$ and $m =m^{(1.4)}(r)$: 
with this we homologously stretch the core as $r\rightarrow r' = (r_c/r_c^{(1.4)}) r$, $\rho(r') = \rho^{(1.4)}(r)$ and $m(r') = (m_c/m_c^{(1.4)}) m^{(1.4)}(r)$.

Our Markov chain Monte Carlo (MCMC) driver \texttt{MXMX} is specially designed to efficiently drive 
\texttt{NSCool} and it applies the basic technics of \texttt{emcee} \citep{foreman-mackey:2013aa}.
It can handle arbitrary numbers of walkers in an arbitrary number of tempered chains, moving either
in individual random walks or in afine invariant stretches \citep{Christen:2007fj,goodman:2010aa}.
For the present work we found the optimal configuration was having 5 tempered chains, with ``temperatures''
$T= 1, 2, 5, 10$, and $100$, each having 100 walkers. 
The basic chain applied stretches and the other chains simple walks.
After initial burn-in, the chains of scenarios ``A'' and ``B'' had more than 2 millions points while
the one of scenario ``C'' had more than four millions points.
We calculated the integrated autocorrelation lengths \citep{Sokal:1997aa}
$\tau$ of each parameter of each walker: typical values are $50-100$ and the longest ones do not exceed 200
in all three scenarios.
We thus have, in each scenario, more than a hundred effectively independent samples from each walker.

\begin{figure}
	\begin{center}
	\includegraphics[width=0.95\textwidth]{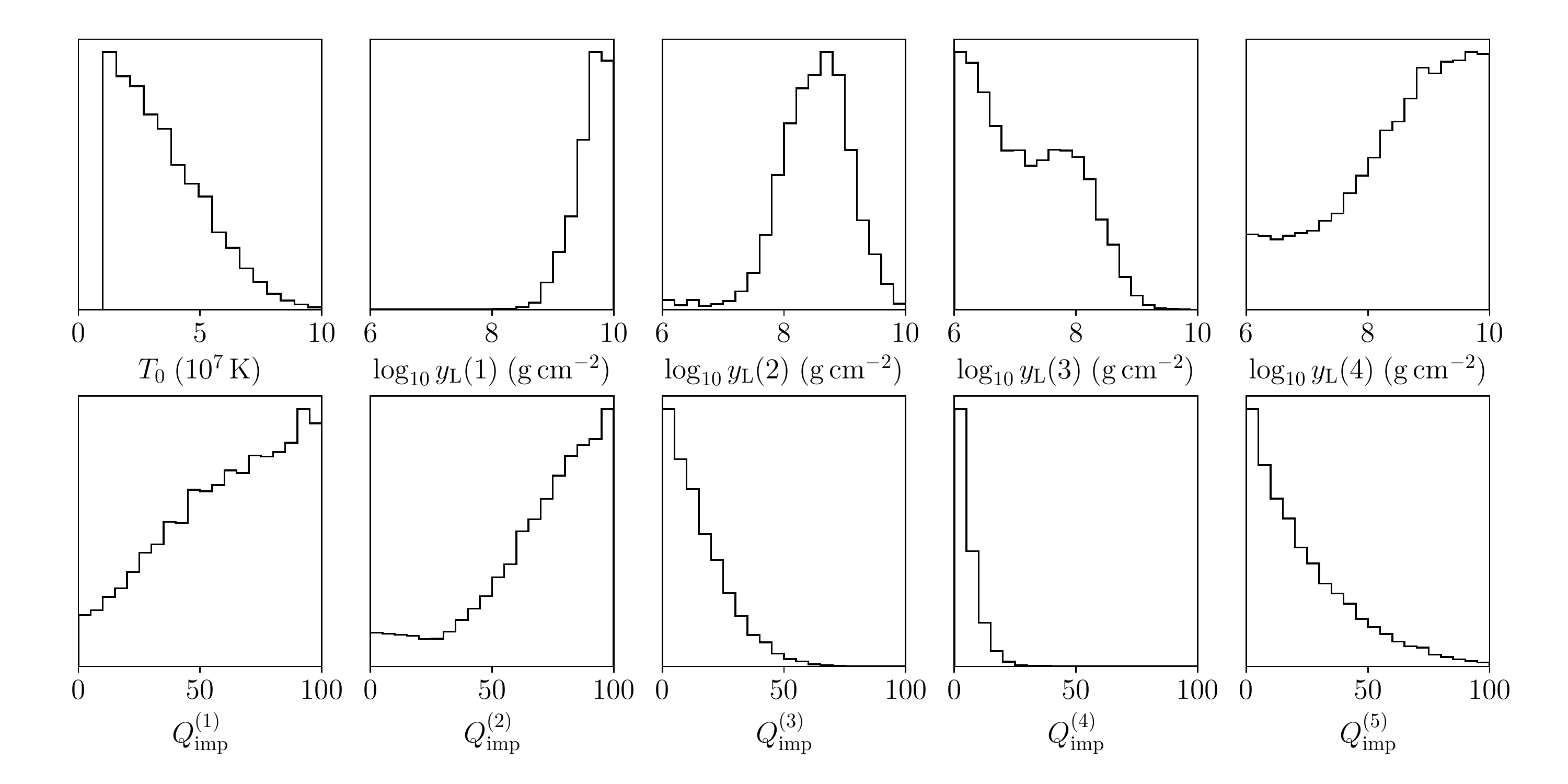}
	\end{center}
	\caption{
	Posterior distribution of the remaining parameters in our scenario ``C'' which are not displayed in 
	\Fig{fig:ScenarioC_Shallow} and \ref{fig:CornerC}.
	}
	\label{fig:Hist2}
\end{figure}

For completeness we display in \Fig{fig:Hist2} the posterior distributions of the remaining 10 parameters of our
scenario ``C'' MCMC that were not displayed in \Fig{fig:ScenarioC_Shallow} and \ref{fig:CornerC} for not being crucial to
the purpose of this work.
Interesting to notice are the preference for cold cores, $T_0$, and the contrast in the distribution of 
$Q_\mathrm{imp}$ in region 1 and 2, i.e., essentially below neutron drip, versus regions 3, 4, 5, i.e., above neutron drip.
One expects a hyperburst to significantly reduce the impurity content as all low $Z$ nuclei are burned into iron peak nuclei while the MCMC prefers high impurity in the region (1), i.e., at densities below $10^{11}$ g cm$^{-3}$, precisely where the hyperbust happened (and in region (2) whose matter should have also been processed by previous hyperbursts).
However, low $Q_\mathrm{imp}$ are {\it not excluded} in regions (1) and (2), only disfavored, and low $Q_\mathrm{imp}$ values actually just favor a thick crust, i.e., not too high neutron star masses and not too small radii.
There is, thus, nothing contradictory, nor conclusive, in these impurity content posteriors. They would favor the secondary peak, low $M$ and large $R$,  in the $M$ and $R$ distributions discussed in \Sec{Sec:hburst}, if one added the new prior that the hyperburst would result in low $Q_\mathrm{imp}$, at least in region 1.
(For this reason we choose our illustrative model for \Fig{fig:Bigbadaboom} and \ref{fig:Bigbadaboom2} as a 1.6 $M_\odot$ star with an 11.2 km radius.)

%------------------------------------------------------------------------------------------------
%------------------------------------------------------------------------------------------------

%------------------------------------------------------------------------------------------------
%------------------------------------------------------------------------------------------------

%%%%%%%%%%%%%%%%%%%%%%%%%%%%%%%%%%%%%%%%%%%%%%%%%%%%%%%%%%%%%%%%%%%%%%%%%%%%%%%%
\section{Electron Captures on $\alpha$-Nuclei}
\label{Sec:ecaptures}
%%%%%%%%%%%%%%%%%%%%%%%%%%%%%%%%%%%%%%%%%%%%%%%%%%%%%%%%%%%%%%%%%%%%%%%%%%%%%%%%

For studying electron captures, we have followed the scheme of \citet{Sato:1979aa} and \citet{Haensel:1990kx}. 
We choose an initial nucleus $(Z,N)$, $N=A-Z$ being the neutron number of the nucleus, and follow its evolution as pressure is increased.
We only consider $\alpha$ nuclei, i.e., $Z$ is even and $Z=N$.
When reaching a critical point where the electron chemical potential $\mu_e$ is large enough a first electron capture occurs resulting in an odd-odd nucleus
$(Z-1,N+1)$ at which point, due to pairing, a second electron capture immediately happens leading to an even-even nucleus  $(Z-2,N+2)$.
The criterion for the first electron capture is that
\begin{equation}
W_N(Z,N) + W_C(Z, n_e) + \mu_e(n_e) = W_N(Z-1,N+1) + W_C(Z-1, n_e)
\label{eq:capture}
\end{equation}
where $W_N(Z,N) = M_N(Z,N) c^2$ is the energy, $M_N$ being its mass, of the nucleus $(Z,N)$, 
$W_C(Z, n_e)$ the Coulomb energy of the nucleus' Wigner-Seitz cell immersed in a medium with electron density $n_e$, 
and $\mu_e$ is the electron chemical potential.
For the Coulomb energy we apply the ion sphere value $W_C(Z, n_e) = -0.9 Z^{5/3} \, e^2/a_e$ where $e$ is the elementary charge unit and
$a_e \equiv (3/4\pi n_e)^{1/3}$ \citep{Shapiro:1986wz}.
We deduce $M_N(Z,N)$ from the atomic mass $M_\mathrm{atom}(Z,N)$ of the AME2020 table \citep{Wang:2021vd} 
as $M_N(Z,N) = M_\mathrm{atom}(Z,N) - Z m_e + B_\mathrm{el}$ where $m_e$ is the electron mass and the electrons binding energy 
is approximated as $B_\mathrm{el} = 14.4381 \, Z^{2.39} + 1.55468\times 10^{-6} \, Z^{5.35} \; \mathrm{eV}$  \citep{2003Lunney}.
We then keep increasing the pressure looking for further electron captures leading to $(Z-4,N+4)$, a.s.o.,
until the neutron drip point is reached.
The resulting evolutions are presented in Fig.~\ref{fig:ecaptures}.

%%%%%%%%%%%%%%%%%%%%%%%%%%%%%%%%%%%%%%%%%%%%%%%%%%%%%%%%%%%%%%%%%%%%%%%%%%%%%%%%
\section{Nuclear Fusion Processes and Screening}
\label{Sec:Nucl}
%%%%%%%%%%%%%%%%%%%%%%%%%%%%%%%%%%%%%%%%%%%%%%%%%%%%%%%%%%%%%%%%%%%%%%%%%%%%%%%

We consider nuclei of charge and mass numbers $Z_i$ and $A_i$ and mass $M_i$ undergoing a
fusion reaction $(A_1,Z_1)+(A_2,Z_2) \to (A_c,Z_c)$. 
For this we need to know the fusion cross section $\sigma$ and calculate the fusion rate
including a proper treatment of plasma screening.
The cross section is commonly expressed as
\begin{equation}
\sigma(E) = \frac{S(E)}{E} \exp(-2\pi \eta)
\label{eq:sigma}
\end{equation}
where $S(E)$ is the ``astrophysical S-factor'', $E$ being the center of mass energy,
and $\eta=Z_1Z_2e^2/\hbar v$ is the Gamow parameter,
with $v=\sqrt{2E/\mu}$ the relative velocity, $\mu = M_1 M_2/(M_1+M_2)$ being the reduced mass.
For our purpose we need an extended set of fusion reactions involving neutron rich nuclei and we employ
the results of \citet{Afanasjev:2012uv} that provide a consistent scheme covering thousand of possible reactions.
These authors calculated $S(E)$ using the S\~{a}o Paulo potential with 
the barrier penetration formalism (SP-BP: \citealt{Chamon:2002we,Gasques:2005tr})
and provide simple analytical fits to their results.

In the purely thermonuclear regime, i.e., at high temperatures where screening can be neglected, 
the fusion rate is given by the standard result (e.g., \citealt{2012Kippenhahn})
\begin{equation}
R = \frac{n_1 n_2}{1 + \delta_{12}} \, S(E^\mathrm{pk}) \, \frac{r_\mathrm{B}}{\hbar} \, P \, F
\label{eq:rate}
\end{equation}
where $n_i$ is the number density of nucleus $i$, $r_\mathrm{B} = \hbar^2/(2\mu Z_1 Z_2 e^2)$, and $S(E)$ is evaluated at the Gamow peak energy $E^\mathrm{pk} = (\pi k_B T/2)^{2/3} E_a^{1/3}$ with $E_a = Z_1 Z_2 e^2/r_\mathrm{B}$.
$F = \exp[-\tau]$ describes the Coulomb barrier penetration with $\tau = 3E^\mathrm{pk}/k_BT$ and the prefactor
$P=16\tau^2/(3^{5/2}\pi)$. (The $1+\delta_{12}$ term avoids double counting in the case nuclei 1 and 2 are identical.)

Based on their previous work \citep{Gasques:2007ur}, \citet{Afanasjev:2012uv} state that typical uncertainties on
their $S(E)$ calculations are of the order of 2 to 4 for stable nuclei but can be about a factor 10 in the cases of
neutron rich nuclei and even up to 100 at low energies.
For example, in the case of the three typical fusion reactions of 
$^{12}$C+$^{12}$C, $^{12}$C+$^{16}$O, and $^{16}$O+$^{16}$O
we find that the thermonuclear rates from Eq.~\ref{eq:rate} using $S(E)$ from \citet{Afanasjev:2012uv}
can be up to 4-5 times smaller that the classical values from \citet{Caughlan:1988aa}.
For another comparison, including important cases of reactions with no experimental measurements, 
\citet{Umar2012} performed dynamical density-constrained time dependent Hartree-Fock (DC-TDHF) calculations of $S(E)$
for several C and O reactions and compared them to the static SP-BP results:
in the cases of $^{12}$C+$^{16}$O, and $^{16}$O+$^{16}$O they obtain excellent agreements with 
experimental values for the DC-TDHF results (while the SP-BP results are about 4 time smaller as seen above),
in the cases of $^{12}$C+$^{24}$O and $^{16}$O+$^{24}$O they find that the SP-BP results are also 
smaller that the DC-TDHF ones, by about one order of magnitud, while in the case of the $^{24}$O+$^{24}$O
the situation reversed with SP-BP values being larger that DC-TDHF ones. 
(No experimental values are available for theses last three reactions.)
As seen below, variations of even a factor 10 in $S(E)$ have very little impact on our results.
Notice that resonances are not included in any of these models that only provide an average $S(E)$ but their
effect is small as long as their width is much smaller than the width of the Gamow peak \citep{Yakovlev:2006wf} .

\begin{figure*}[t]
	\begin{center}
	\includegraphics[width=0.99\textwidth]{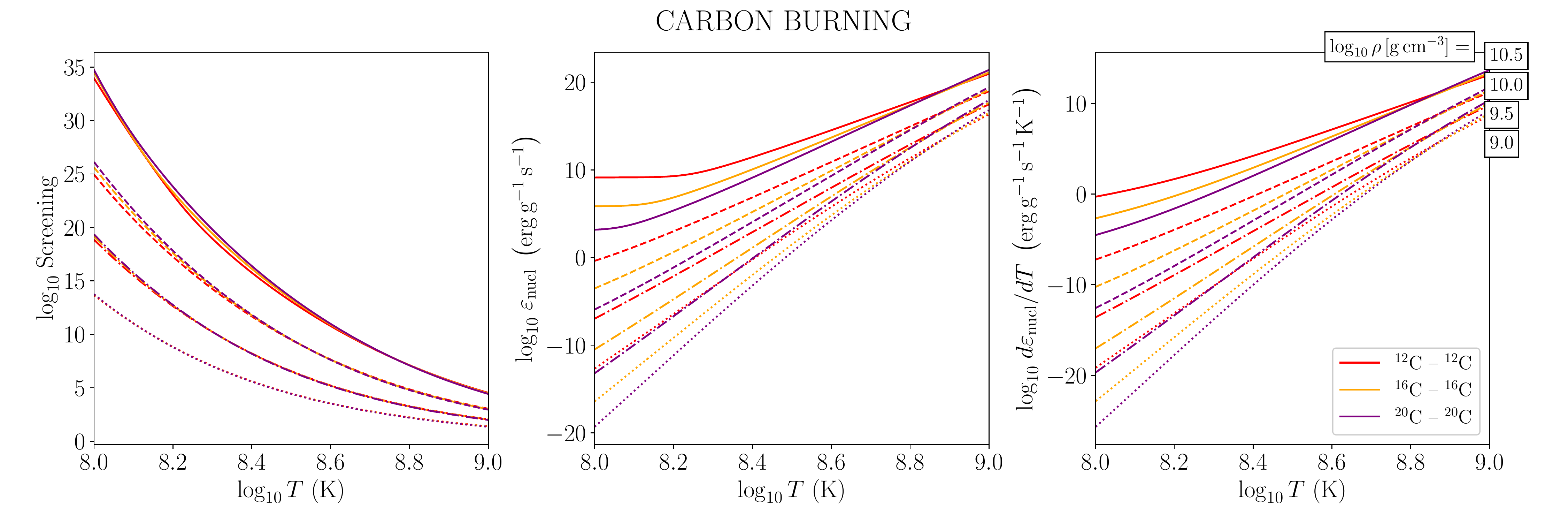}
	\includegraphics[width=0.99\textwidth]{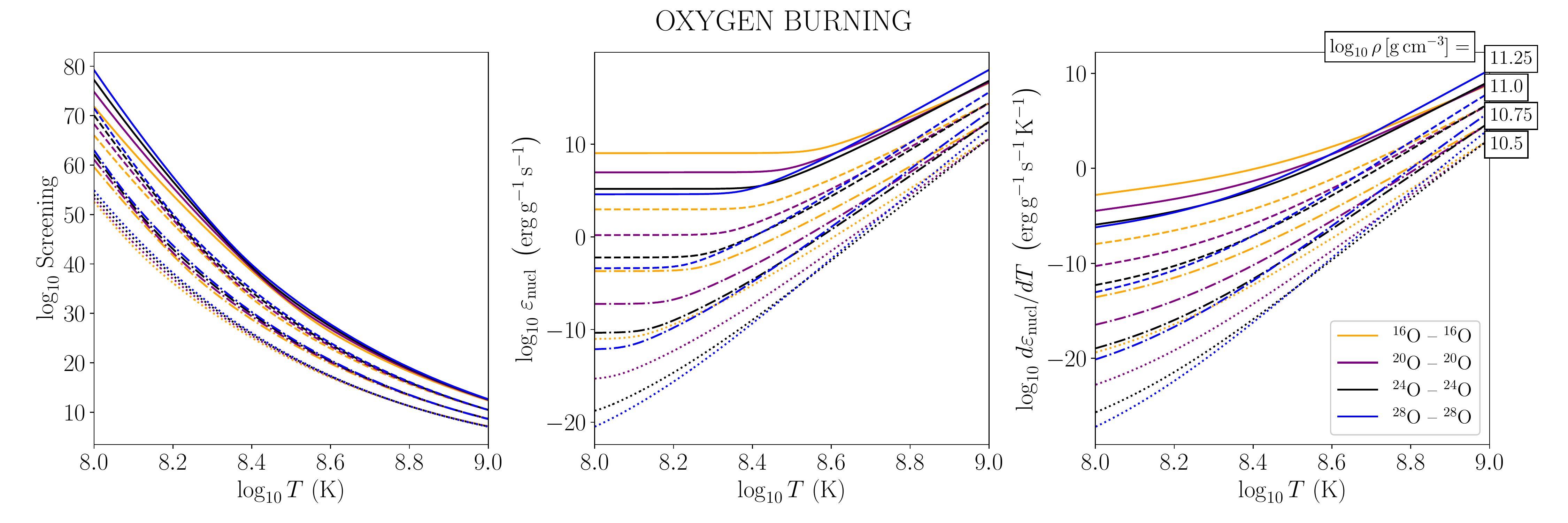}
	\includegraphics[width=0.99\textwidth]{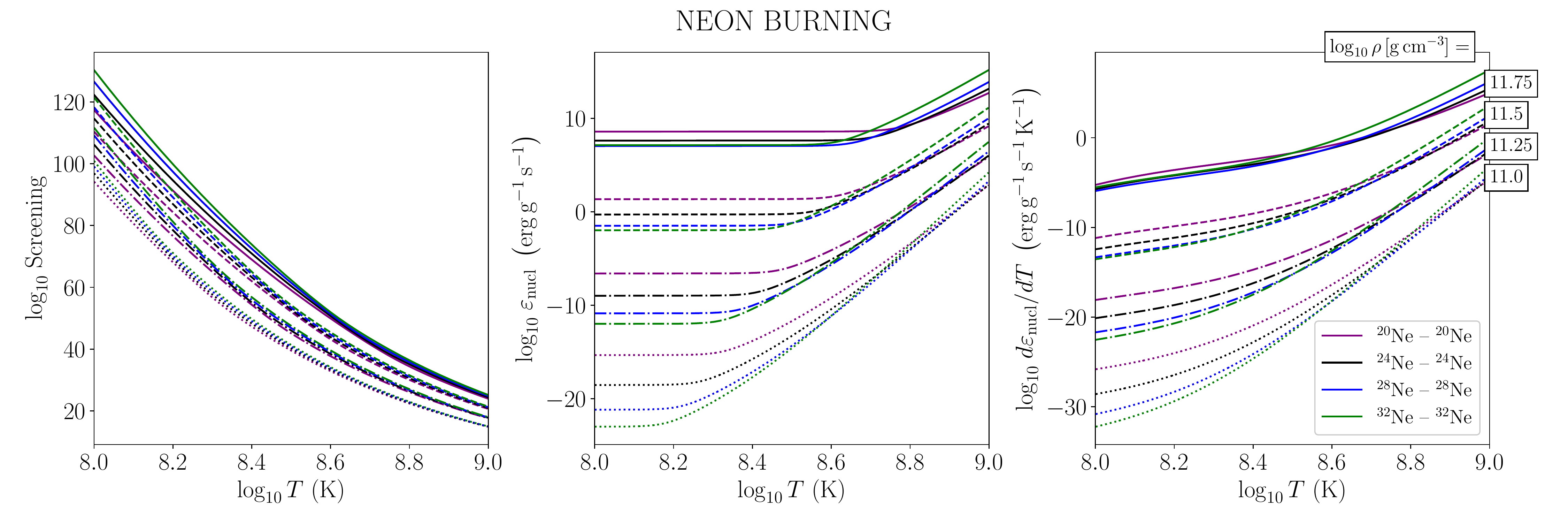}
	\end{center}
	\caption{Screening factors, energy generation rates, $\varepsilon_\mathrm{nucl}$, and their temperature sensitivity, 
	$d\varepsilon_\mathrm{nucl} / dT$, for fusions of several isotopes of C, O, and Ne, at chosen densities.
	We used the ``optimal'' reaction rates and in all cases assumed a mass fraction $X = 10^{-3}$ within a background of $^{56}$Ca.
	}
	\label{fig:burning}
\end{figure*}

The second ingredient needed to calculate realistic fusion rates in the dense medium of the neutron star
crust is proper inclusion of the screening of the Coulomb repulsion. 
In our cases this effect increases the rates by {\it many tens} of orders of magnitude.
We employ the treatment proposed by \citet{Gasques:2005tr} and \citet{Yakovlev:2006wf} 
which allows to cover the whole range from the $T=0$ pycnonuclear regime up to the thermonuclear one 
by adjusting $E^\mathrm{pk}$ and appropriately modifying the two functions $P$ and $F$ in \Eq{eq:rate}.
In the thermonuclear regime screening can be accurately incorporated while in the opposite pycnonuclear regime
uncertainties are enormous, possibly up to 10 orders of magnitude.
We will apply the three representative case of \citet{Yakovlev:2006wf} that give minimal and maximal screening
effects as well at their optimal model.
From the rate per unit volume $R$ one deduces the energy generation rate, per unit mass, $\varepsilon_\mathrm{nucl} = QR/\rho$ where $Q$ is the energy
released per reaction.
Representative $Q$-values are listed in Table~\ref{tb:Q}.
Relevant screening factors, energy generation rates $\varepsilon_\mathrm{nucl}$ and their temperature sensitivity $d\varepsilon_\mathrm{nucl} / dT$
are plotted in \Fig{fig:burning}.
The flat portion of the $\varepsilon_\mathrm{nucl}$ curve exhibit the pycnonuclear regime and comparison with \Fig{fig:fusion}
show that the triggering of the hyperburst occurred in the regime of transition from purely pycnonuclear to thermonuclear.

%------------------------------------------------------------------------------------------------
\begin{table}
\caption{Energy released per reaction, $Q$, and per nucleon, $Q/A$, both in MeV, for fusions between nuclei
with same charge $Z$ and mass numbers $A_1$ and $A_2$.
Nuclear binding energies are taken from the 2020 Atomic Mass Evaluation \citep{Huang:2021vb,Wang:2021vd}.
}
\begin{center}
\begin{tabular}{ccccc|ccccc|ccccc}
\hline
$Z_1=Z_2$ & $A_1$ & $A_2$ & $Q$ & $Q/A$ & $Z_1=Z_2$ & $A_1$ & $A_2$ & $Q$ & $Q/A$ & $Z_1=Z_2$ & $A_1$ & $A_2$ & $Q$ & $Q/A$ \\
\hline
 6 & 12 & 12 & 13.9 &  0.58   &    8 & 16 & 16 & 16.5   &  0.52   &   10 & 20 & 20 & 20.8 &  0.52 \\
 6 & 12 & 16 & 28.7 &  1.03   &    8 & 16 & 20 & 29.7   &  0.83   &   10 & 20 & 24 & 28.5 &  0.65 \\
 6 & 12 & 20 & 38.3 &  1.20   &    8 & 16 & 24 & 36.6   &  0.92   &   10 & 20 & 28 & 48.5 &  1.01 \\
 6 & 16 & 16 & 28.2 &  0.88   &    8 & 16 & 28 & 56.5   &  1.29   &   10 & 20 & 32 & 64.2 &  1.24 \\
 6 & 16 & 20 & 30.8 &  0.86   &    8 & 20 & 20 & 30.4   &  0.76   &   10 & 24 & 24 & 32.3 &  0.67 \\
 6 & 20 & 20 & 25.5 &  0.64   &    8 & 20 & 24 & 31.5.  &  0.72   &   10 & 24 & 28 & 39.6 &  0.76 \\
    &      &      &         &            &    8 & 20 & 28 & 43.5.  &  0.91   &   10 & 24 & 32 & 44.6 &  0.80 \\
    &      &      &         &            &    8 & 24 & 24 & 24.6   &  0.51   &   10 & 28 & 28 & 36.1 &  0.64 \\
    &      &      &         &            &    8 & 24 & 28 & 70.6   &  1.36   &   10 & 28 & 32 & 37.3 &  0.62 \\
    &      &      &         &            &    8 & 28 & 28 & 104.2 &  1.86   &   10 & 32 & 32 & 74.0 &  1.16 \\
\hline
\end{tabular}
\end{center}
\label{tb:Q}
\end{table}
%------------------------------------------------------------------------------------------------

%------------------------------------------------------------------------------------------------
%------------------------------------------------------------------------------------------------

%%%%%%%%%%%%%%%%%%%%%%%%%%%%%%%%%%%%%%%%%%%%%%%%%%%%%%%%%%%%%%%%%%%%%%%%%%%%%%%%
%%%%%%%%%%%%%%%%%%%%%%%%%%%%%%%%%%%%%%%%%%%%%%%%%%%%%%%%%%%%%%%%%%%%%%%%%%%%%%%%
%\bibliography{MAXI_2021}{}
\bibliographystyle{aasjournal}
%%%%%%%%%%%%%%%%%%%%%%%%%%%%%%%%%%%%%%%%%%%%%%%%%%%%%%%%%%%%%%%%%%%%%%%%%%%%%%%%
%%%%%%%%%%%%%%%%%%%%%%%%%%%%%%%%%%%%%%%%%%%%%%%%%%%%%%%%%%%%%%%%%%%%%%%%%%%%%%%%

%%%%%%%%%%%%%%%%%%%%%%%%%%%%%%%%%%%%%%%%%%%%%%%%%%%%%%%%%%%%%%%%%%%%%%%%%%%%%%%%
%%%%%%%%%%%%%%%%%%%%%%%%%%%%%%%%%%%%%%%%%%%%%%%%%%%%%%%%%%%%%%%%%%%%%%%%%%%%%%%%
%%%%%%%%%%%%%%%%%%%%%%%%%%%%%%%%%%%%%%%%%%%%%%%%%%%%%%%%%%%%%%%%%%%%%%%%%%%%%%%%
%%%%%%%%%%%%%%%%%%%%%%%%%%%%%%%%%%%%%%%%%%%%%%%%%%%%%%%%%%%%%%%%%%%%%%%%%%%%%%%%
%%%%%%%%%%%%%%%%%%%%%%%%%%%%%%%%%%%%%%%%%%%%%%%%%%%%%%%%%%%%%%%%%%%%%%%%%%%%%%%%
\end{document}